\documentclass[aps,pre,showpacs,reprint,superscriptaddress,amsmath,amsfonts]{revtex4-1}
\usepackage{amssymb}
\usepackage{graphicx}
\allowdisplaybreaks
\usepackage{pgfplots}
\usepackage{url}

\usepackage{color}

\DeclareMathOperator\smol{\Omega}
\newcommand*\kB{\ensuremath{k_\mathrm{B}}}
\DeclareMathOperator\Real{Re}
\DeclareMathOperator\Imag{Im}
\newcommand{\A}{\tensor{A}}
\newcommand{\X}{{\scriptscriptstyle{X}}}
\newcommand{\V}{{\scriptscriptstyle{V}}}
\newcommand{\+}{{\scriptscriptstyle{+}}}
\newcommand{\pr}{\partial}
\renewcommand{\tensor}[1]{\boldsymbol{#1}}
\renewcommand{\vec}[1]{\boldsymbol{#1}}
\newcommand\rem[1]{\textbf{#1}}

\newcommand{\caA}{{\mathcal A}}
\newcommand{\caB}{{\mathcal B}}
\newcommand{\caC}{{\mathcal C}}
\newcommand{\caD}{{\mathcal D}}
\newcommand{\caE}{{\mathcal E}}
\newcommand{\caF}{{\mathcal F}}
\newcommand{\caG}{{\mathcal G}}
\newcommand{\caH}{{\mathcal H}}
\newcommand{\caI}{{\mathcal I}}
\newcommand{\caJ}{{\mathcal J}}
\newcommand{\caK}{{\mathcal K}}
\newcommand{\caL}{{\mathcal L}}
\newcommand{\caM}{{\mathcal M}}
\newcommand{\caN}{{\mathcal N}}
\newcommand{\caO}{{\mathcal O}}
\newcommand{\caP}{{\mathcal P}}
\newcommand{\caQ}{{\mathcal Q}}
\newcommand{\caR}{{\mathcal R}}
\newcommand{\caS}{{\mathcal S}}
\newcommand{\caT}{{\mathcal T}}
\newcommand{\caU}{{\mathcal U}}
\newcommand{\caV}{{\mathcal V}}
\newcommand{\caW}{{\mathcal W}}
\newcommand{\caX}{{\mathcal X}}
\newcommand{\caY}{{\mathcal Y}}
\newcommand{\caZ}{{\mathcal Z}}
\newcommand{\om}{{\omega}}
\newcommand{\omb}{{\bar\omega}}
\newcommand{\tom}{{\tilde\omega}}
\newcommand{\diff}{\text{d}}
\newcommand\avg[1]{\langle #1\rangle}

\newcommand{\mean}[1]{{\left< #1 \right>}}

\begin{document}

\title{Non-Isothermal Fluctuation-Dissipation Relations and Brownian Thermometry}

\newcommand\mpi{\affiliation{Max Planck Institute for Mathematics in the Sciences, Inselstr. 22,
04103 Leipzig, Germany}}
\newcommand\ul{\affiliation{Institut f\"ur Theoretische Physik, Universit\"at Leipzig,  Postfach 100 920, D-04009 Leipzig, Germany}}
\newcommand\ule{\affiliation{Department of Physics and Astronomy, University of Leeds, LS2 9JT Leeds, England}}
\author{G.~Falasco}\ul
\author{M.~V.~Gnann}\mpi
\author{K.~Kroy}\ul

\begin{abstract}
  The classical theory of Brownian motion rests on fundamental laws of
  statistical mechanics, such as the equipartition theorem and the
  fluctuation-dissipation theorem, which are not applicable in
  non-isothermal situations. We derive the generalized
  fluctuation-dissipation relations and Langevin equations governing
  such non-isothermal Brownian motion, including explicit results for
  the frequency-dependent noise temperature and Brownian thermometry
  far from equilibrium.
\end{abstract}

\maketitle

The perpetual thermal dance of the molecules of a fluid manifests
itself in incessant erratic movements of small suspended particles. If
a quantitative characterization of this mesoscopic motion is
available, it can serve as a Brownian thermometer. More than a century
ago, A. Einstein  \cite{einstein.1905} provided such a mesoscopic stochastic model demonstrating that the
3-dimensional position and velocity fluctuations of isothermal
Brownian motion are indeed given by the thermal energy $\kB T$,
\begin{equation}\label{eq:einstein}
  \begin{split}
    \underset{t\to\infty}\lim \avg{[\vec X(t)- \vec X(0)]^2}/6t &
    = 3\kB T/\zeta\\  
     \underset{t\to 0}\lim \avg{\vec V(t) \cdot\vec V(0)} & = 3\kB
     T/M \,,
  \end{split}
\end{equation}
filtered through the friction $\zeta$ and the renormalized
 inertial mass $M$ \cite{Landau.1987} of the particle, respectively.  Note that
both coefficients characterize the fluid-particle interactions, and
that the Brownian thermometer not only measures $T$ but additionally
witnesses the atomistic origin of the fluctuations through the
appearence of Boltzmann's constant (originally the gas constant divided
by the Loschmidt number).  The experimental verification
of  Eqs.~(\ref{eq:einstein}) could thus be interpreted as a confirmation of
the atomistic structure of matter \cite{perrin.1926}, and their
generalization in the form of the fluctuation-dissipation theorem
(FDT) became a corner stone of statistical mechanics
\cite{Kubo.1992}.  
A more accurate quantum-mechanical description modifies Eqs.~(\ref{eq:einstein}),
showing that eigenmodes of eigenfrequencies $\omega \gg \kB T/\hbar$
freeze out exponentially \cite{callen.1951}. The physical interpretation of the quantum
version of the FDT is then that a Brownian thermometer probes a
universal energy spectrum \footnote{In
  fact, the formula is not completely general but assumes a bosonic
  heat bath.}  $\hbar\omega\coth (\hbar\omega/2\kB T)$ rather
than a universal thermal energy $\kB T$.

In this Letter, we show that a very similar (though somewhat less
universal) situation is encountered far from equilibrium. We
generalize Eqs.~(\ref{eq:einstein}) and the FDT to the case of a
classical Brownian particle in a non-isothermal fluid by explicitly
calculating the appropriate energy spectrum $\kB\caT(\omega)$ of the
thermal noise. Thereby, we extend Einstein's theory of a Brownian
thermometer to non-isothermal solvents.  Similar as the kinetic
coefficients $\zeta$ and $M$, and unlike the universal quantum
spectrum, $\caT(\omega)$ is found to be a non-universal function of
the particle-fluid interactions. This should not come as a
surprise. Due to the lack of a zero'th law, the reading of any
thermometer operating far from equilibrium should certainly be
expected to ``depend on its orientation, shape, surface properties
(...) and other variables in the system being observed"
\cite{mclennan.1989, casaz.2003}.  Nevertheless, our derivation, much
like its equilibrium counterpart, establishes a comparatively simple,
yet precise, mesoscopic model of Brownian motion that is essentially
independent of molecular details. Also like in equilibrium, it remains a paradigm for
fluctuations in other mesoscopic devices operating under
non-isothermal conditions, e.g.\ in electrical engineering (Johnson--Nyquist noise)
and nanophotonics (antenna noise).

\emph{General theory}---Brownian motion in a thermally inhomogeneous bath is a vivid research subject \cite{Matsuo.2000, Bringuier.2007, Celani.2012, Polettini.2013}. Yet, the dynamical equations on which most of this work hinges are often taken for granted. Explicit derivations \cite{perez.1994, zubarev.1968, shea.1996} assume the particle to be sensitive only to local (in space and time) thermal fluctuations, conforming with the principles of non-equilibrium thermodynamics \cite{DeGroot}. It is rarely acknowledged \cite{VanKampen.1986} that such an assumption disregards the solvent dynamics, namely how the momentum spreads from the particle into the fluid and is returned
in the form of colored noise. This issue has recently received
considerable attention for isothermal Brownian motion
\cite{Franosch.2011, Kheifets.2014} and is our main focus,
here. We take the solvent to be a simple liquid described by the theory of fluctuating hydrodynamics
\cite{Sengers.2006}\footnote{Note that our discussion therefore includes
  ``kinematic'' thermophoresis due to the inhomogeneous noise
  strength, while it does not account for ``molecular'' thermophoresis
  due to molecular interactions.}.  We assume it to be incompressible, locally equilibrated at
temperature $T(\vec r)$, and mechanically coupled to a
colloidal sphere of radius $R$ at position $\vec X$ via standard 
boundary conditions (no influx, no slip).  Contracting this description to an equation of
motion for the colloidal particle, alone, proceeds along similar lines
as in global equilibrium \cite{Hauge.1973, Bedaux.1974}, but requires a number of additional
considerations. In the following, we summarize the somewhat lengthy
and formal procedure in an intuitive, but essentially correct way and
refer the interested reader to Refs.~\cite{SI, Falasco.2014} for more
details.

The key point is that, due to momentum conservation, the solvent as a whole
is responsible for the forces acting on the Brownian particle. In the Stokes limit, momentum
is redistributed by vorticity diffusion, the diffusivity being given
by the local kinematic solvent viscosity $\nu$. The characteristic
rate of momentum equilibration between particle and solvent is thus
$\omega_\text{f} = 2\nu/R^2$.  Depending on the frequency $\omega$ of
its motion, the colloidal particle accordingly exchanges momentum with
a small ($\omega \gg \omega_\text{f}$) or large ($\omega \ll
\omega_\text{f}$) solvent volume.
In other words, there exists a frequency-dependent coupling, mediated
by the hydrodynamic modes, between the particle and distant volume
elements $\Delta \caV$ of the solvent.   In a discretized
representation, we enumerate these volume elements by $\alpha$, instead
of $\vec r$. The particle-solvent coupling can then be fully encoded in the local
response function $\phi_{ij}^\alpha(\vec X,\omega)$  \footnote{See \cite{SI} for its formal expression, and
  \cite{Falasco.2014} for an explicit approximate formula.} of
the fluid momentum (at position $\alpha$) in the $i-$direction to the
particle motion (at position $\vec X$) in the $j-$direction. Since
$\phi_{ij}^\alpha(\vec X,\omega)$ 
quantifies how kinetic energy is turned into heat at the local reservoirs
$\alpha$,  maintained at temperatures $T^\alpha$,
it amounts to a local frequency-dependent friction tensor. 
The reverse process of turning heat into kinetic energy gives rise to random
fluctuations of the particle velocity, which we represent by a
stochastic force $\xi_i^\alpha(\vec X,t)$. By virtue of the assumption of local
thermal equilibrium, this force is Gaussian with zero mean, and its strength is
uniquely determined by the local temperature $T^\alpha$ and the
(Fourier-transformed) response $\phi_{ij}^\alpha(\vec X,t)$ through the local FDT,
\begin{equation*}
  \langle \xi_i^\alpha(\vec X,t)  \xi_j^\beta(\vec X,t') \rangle \approx
  T^\alpha \phi_{ij}^\alpha(\vec X,t-t')  \Delta\caV \delta_{\alpha \beta}\,.
\end{equation*}
The total acceleration of the Brownian particle of mass $m$ 
can accordingly be written as a sum over the contributions from all volume elements,
\begin{equation*}
m \ddot X_i \approx  \sum_{\alpha}\left[- \int_{-\infty}^t\!
  \sum_{j=1}^3\phi_{ij}^\alpha(\vec X, t - t^\prime) \Delta\caV  \dot
  X_j(t^\prime) \diff t^\prime + \xi^{\alpha}_i \right] \,.
\end{equation*}
In the continuum limit, $\Delta \caV\to0$, the terms in the sum can be
added up to the friction and noise terms
\begin{equation*}
\zeta_{ij}(\vec X, t)\approx \sum_{\alpha} \phi_{ij}^\alpha(\vec X,t)\Delta \caV
\end{equation*}
and $\xi_i(\vec X,t) \approx \sum_{\alpha} \xi_i^{\alpha}(\vec X,t)$, respectively. Thereby,
the equation of motion can be rewritten in the form of the generalized
Langevin equation
\begin{equation}\label{GLE}
m \ddot {\vec X}(t) = -\int_{-\infty}^{t} \vec \zeta(\vec X, t-t')\cdot  \dot
{\vec X}(t')\diff t'  + \vec{\xi}(\vec X, t)
\end{equation}
(plus an optional external force term). The non-equilibrium Gaussian
noise $\vec{\xi}$ has vanishing
mean, and its correlations in the frequency domain read:
\begin{equation}\label{noise}
  \langle \xi_i(\vec X, \omega) \xi_j(\vec X,\omega')\rangle =
  \kB \mathcal{T}_{ij}(\vec X, \omega) \zeta_{ij}(\vec X, \omega) \delta(\omega+\omega')\,.
\end{equation}
In contrast to the familiar isothermal situation \cite{Franosch.2011},
the color of the Brownian noise is not governed by the
frequency-dependence of the friction alone, but also by that of the
tensorial \emph{noise temperature}
\begin{equation}\label{Tdef}
\mathcal{T}_{ij}(\vec X, \omega)=\frac{\int_{\caV}  \phi_{ij}(\vec X,\vec r,
  \omega) T(\vec r) \, \diff \vec r}{\int_{\caV}
  \phi_{ij}(\vec X,\vec r, \omega)\, \diff \vec r}\,.
\end{equation}

Equations (\ref{GLE}-\ref{Tdef}) constitute the central results of our
contribution. In particular, the last two equations provide a
generalization of the FDT (of the second kind) for a Brownian particle
in a non-isothermal solvent.  Due to the
long-range hydrodynamic interactions between the particle and the
fluid the noise temperature
is manifestly non-local---Eq.~(\ref{Tdef}) has the form of a
spatial average over the local fluid temperature $T(\vec r)$ around the
instantaneous particle position $\vec X$.  The weighting by the response tensor $\phi_{ij}$  
reveals that $\mathcal{T}_{ij}$ is not a property of the solvent or the particle, alone, but
characterizes their mutual coupling. Via the symmetry of $\phi_{ij}$
in $\omega$ \cite{Hauge.1973}, it inherits the time-reversal
invariance of the microscopic dynamics of the bath. And via the
spatial dependence of $\phi_{ij}$ it ``knows'' about the particle radius, boundary
conditions, and tranport mode (translation or rotation) \cite{Falasco.2014, Rings.2012}.
The tensorial structure of $\mathcal{T}_{ij}$ reflects that, for an
arbitrary temperature field $T(\vec r)$, hydrodynamic modes may carry
different amounts of thermal energy along different spatial
directions. If one interprets the colloidal particle as a Brownian
thermometer sampling the solvent temperature field $T(\vec r)$ via
the mesoscopic intermediate $\mathcal{T}_{ij}$, all this should  
be recognized as an important necessary feature rather than a
deficiency. Indeed, the anisotropy and the variation
of $\mathcal{T}_{ij}(\vec X, \omega)$ with its arguments and the
transport mode can serve as measures for the departure from thermal 
equilibrium \cite{Hohenberg.1989}.

Intuitively, the meaning of the noise temperature can probably
be best understood from the observation that it maps the local equilibrium condition of the hydrodynamic model onto a local equilibrium condition in the contracted system, introducing a quantum-type thermal spectrum.  
Due to the underlying local-equilibrium assumption, $\kB
\caT_{ii}(\vec X, \omega)$ literally represents the thermal energy that an
equilibrium bath would have to supply via the noise component $\xi_i$ in order to mimic the effect of the non-isothermal
solvent. That this energy has, in general, a
colored spectrum and depends on the current position of the particle,
and the direction of its motion, is the price one pays for integrating
out the slow non-equilibrium degrees of freedom of the bath. It is worthwhile
pointing out, though, that these ``imperfections'' of the equilibrium
analogy are (partially) mended in certain practically important limits to be discussed below.

\begin{figure}[t]
\begin{tikzpicture}
\begin{axis}[name=mainplot, xmin=0,xmax=0.6,ymin=0 ,ymax=0.5, axis equal image, xlabel=$D(t)$,
ylabel=$\mu(t)$] 
\addplot[color=red,dash pattern=on 4pt off 1pt on 4pt off 4pt ] {x/(1+0.781371)};
\addplot[color=blue,dash pattern=on 4pt off 1pt on 1pt off 1 pt ] {x/(1+5/12)-0.038};
\addplot[color=black] coordinates {
(0,0) (0.002, 0.00112837)(0.003, 0.00169318)(0.004, 0.00225842) 
(0.005, 0.00282407)(0.006, 0.00339013)(0.007, 0.00395662) 
(0.008, 0.00452351)(0.009, 0.00509081)(0.01, 0.00565853)(0.011, 
0.00622665)(0.012, 0.00679518)(0.013, 0.00736412)(0.014, 
0.00793346)(0.015, 0.0085032)(0.016, 0.00907335)(0.017, 
0.00964389)(0.018, 0.0102148)(0.019, 0.0107862)(0.02, 
0.0113579)(0.021, 0.01193)(0.022, 0.0125026)(0.023, 0.0130755) 
(0.024, 0.0136488)(0.025, 0.0142225)(0.026, 0.0147966)(0.027, 
0.015371)(0.028, 0.0159459)(0.029, 0.0165211)(0.03, 0.0170968) 
(0.031, 0.0176728)(0.032, 0.0182492)(0.033, 0.0188259)(0.034, 
0.0194031)(0.035, 0.0199806)(0.036, 0.0205585)(0.037, 
0.0211367)(0.038, 0.0217154)(0.039, 0.0222944)(0.04, 
0.0228738)(0.041, 0.0234535)(0.042, 0.0240336)(0.043, 
0.0246141)(0.044, 0.025195)(0.045, 0.0257762)(0.046, 
0.0263577)(0.047, 0.0269397)(0.048, 0.0275219)(0.049, 
0.0281046)(0.05, 0.0286876)(0.051, 0.029271)(0.052, 0.0298547) 
(0.053, 0.0304388)(0.054, 0.0310232)(0.055, 0.0316079)(0.056, 
0.0321931)(0.057, 0.0327785)(0.058, 0.0333644)(0.059, 
0.0339505)(0.06, 0.034537)(0.061, 0.0351239)(0.062, 0.0357111) 
(0.063, 0.0362986)(0.064, 0.0368865)(0.065, 0.0374747)(0.066, 
0.0380633)(0.067, 0.0386522)(0.068, 0.0392414)(0.069, 
0.039831)(0.07, 0.0404208)(0.071, 0.0410111)(0.072, 0.0416016) 
(0.073, 0.0421925)(0.074, 0.0427837)(0.075, 0.0433753)(0.076, 
0.0439672)(0.077, 0.0445594)(0.078, 0.0451519)(0.079, 
0.0457447)(0.08, 0.0463379)(0.081, 0.0469314)(0.082, 
0.0475252)(0.083, 0.0481193)(0.084, 0.0487138)(0.085, 
0.0493085)(0.086, 0.0499036)(0.087, 0.050499)(0.088, 
0.0510947)(0.089, 0.0516907)(0.09, 0.052287)(0.091, 0.0528837) 
(0.092, 0.0534806)(0.093, 0.0540779)(0.094, 0.0546755)(0.095, 
0.0552733)(0.096, 0.0558715)(0.097, 0.05647)(0.098, 0.0570688) 
(0.099, 0.0576678)(0.1, 0.0582672)(0.101, 0.0588669)(0.102, 
0.0594669)(0.103, 0.0600672)(0.104, 0.0606678)(0.105, 
0.0612686)(0.106, 0.0618698)(0.107, 0.0624713)(0.108, 
0.063073)(0.109, 0.0636751)(0.11, 0.0642774)(0.111, 0.0648801) 
(0.112, 0.065483)(0.113, 0.0660862)(0.114, 0.0666897)(0.115, 
0.0672935)(0.116, 0.0678975)(0.117, 0.0685019)(0.118, 
0.0691065)(0.119, 0.0697114)(0.12, 0.0703166)(0.121, 
0.0709221)(0.122, 0.0715279)(0.123, 0.0721339)(0.124, 
0.0727402)(0.125, 0.0733468)(0.126, 0.0739537)(0.127, 
0.0745608)(0.128, 0.0751683)(0.129, 0.075776)(0.13, 0.0763839) 
(0.131, 0.0769922)(0.132, 0.0776007)(0.133, 0.0782095)(0.134, 
0.0788185)(0.135, 0.0794278)(0.136, 0.0800374)(0.137, 
0.0806473)(0.138, 0.0812574)(0.139, 0.0818678)(0.14, 
0.0824784)(0.141, 0.0830893)(0.142, 0.0837005)(0.143, 
0.0843119)(0.144, 0.0849236)(0.145, 0.0855356)(0.146, 
0.0861478)(0.147, 0.0867603)(0.148, 0.087373)(0.149, 0.087986) 
(0.15, 0.0885992)(0.151, 0.0892127)(0.152, 0.0898265)(0.153, 
0.0904405)(0.154, 0.0910548)(0.155, 0.0916693)(0.156, 
0.092284)(0.157, 0.092899)(0.158, 0.0935143)(0.159, 0.0941298) 
(0.16, 0.0947456)(0.161, 0.0953616)(0.162, 0.0959778)(0.163, 
0.0965943)(0.164, 0.0972111)(0.165, 0.0978281)(0.166, 
0.0984453)(0.167, 0.0990628)(0.168, 0.0996805)(0.169, 
0.100298)(0.17, 0.100917)(0.171, 0.101535)(0.172, 0.102154) 
(0.173, 0.102773)(0.174, 0.103392)(0.175, 0.104011)(0.176, 
0.104631)(0.177, 0.105251)(0.178, 0.105871)(0.179, 0.106491) 
(0.18, 0.107112)(0.181, 0.107733)(0.182, 0.108354)(0.183, 
0.108975)(0.184, 0.109596)(0.185, 0.110218)(0.186, 0.11084) 
(0.187, 0.111462)(0.188, 0.112085)(0.189, 0.112707)(0.19, 
0.11333)(0.191, 0.113953)(0.192, 0.114577)(0.193, 0.1152) 
(0.194, 0.115824)(0.195, 0.116448)(0.196, 0.117072)(0.197, 
0.117697)(0.198, 0.118322)(0.199, 0.118946)(0.2, 0.119572) 
(0.201, 0.120197)(0.202, 0.120823)(0.203, 0.121448)(0.204, 
0.122074)(0.205, 0.122701)(0.206, 0.123327)(0.207, 0.123954) 
(0.208, 0.124581)(0.209, 0.125208)(0.21, 0.125835)(0.211, 
0.126463)(0.212, 0.12709)(0.213, 0.127718)(0.214, 0.128346) 
(0.215, 0.128975)(0.216, 0.129603)(0.217, 0.130232)(0.218, 
0.130861)(0.219, 0.131491)(0.22, 0.13212)(0.221, 0.13275) 
(0.222, 0.13338)(0.223, 0.13401)(0.224, 0.13464)(0.225, 
0.135271)(0.226, 0.135901)(0.227, 0.136532)(0.228, 0.137163) 
(0.229, 0.137795)(0.23, 0.138426)(0.231, 0.139058)(0.232, 
0.13969)(0.233, 0.140322)(0.234, 0.140954)(0.235, 0.141587) 
(0.236, 0.14222)(0.237, 0.142853)(0.238, 0.143486)(0.239, 
0.144119)(0.24, 0.144753)(0.241, 0.145387)(0.242, 0.146021) 
(0.243, 0.146655)(0.244, 0.147289)(0.245, 0.147924)(0.246, 
0.148559)(0.247, 0.149194)(0.248, 0.149829)(0.249, 0.150464) 
(0.25, 0.1511)(0.251, 0.151736)(0.252, 0.152372)(0.253, 
0.153008)(0.254, 0.153644)(0.255, 0.154281)(0.256, 0.154918) 
(0.257, 0.155555)(0.258, 0.156192)(0.259, 0.156829)(0.26, 
0.157467)(0.261, 0.158105)(0.262, 0.158743)(0.263, 0.159381) 
(0.264, 0.160019)(0.265, 0.160658)(0.266, 0.161297)(0.267, 
0.161935)(0.268, 0.162575)(0.269, 0.163214)(0.27, 0.163853) 
(0.271, 0.164493)(0.272, 0.165133)(0.273, 0.165773)(0.274, 
0.166413)(0.275, 0.167054)(0.276, 0.167695)(0.277, 0.168335) 
(0.278, 0.168977)(0.279, 0.169618)(0.28, 0.170259)(0.281, 
0.170901)(0.282, 0.171543)(0.283, 0.172185)(0.284, 0.172827) 
(0.285, 0.173469)(0.286, 0.174112)(0.287, 0.174755)(0.288, 
0.175398)(0.289, 0.176041)(0.29, 0.176684)(0.291, 0.177328) 
(0.292, 0.177971)(0.293, 0.178615)(0.294, 0.179259)(0.295, 
0.179904)(0.296, 0.180548)(0.297, 0.181193)(0.298, 0.181838) 
(0.299, 0.182483)(0.3, 0.183128)(0.301, 0.183773)(0.302, 
0.184419)(0.303, 0.185065)(0.304, 0.18571)(0.305, 0.186357) 
(0.306, 0.187003)(0.307, 0.187649)(0.308, 0.188296)(0.309, 
0.188943)(0.31, 0.18959)(0.311, 0.190237)(0.312, 0.190885) 
(0.313, 0.191533)(0.314, 0.19218)(0.315, 0.192828)(0.316, 
0.193477)(0.317, 0.194125)(0.318, 0.194774)(0.319, 0.195422) 
(0.32, 0.196071)(0.321, 0.19672)(0.322, 0.19737)(0.323, 
0.198019)(0.324, 0.198669)(0.325, 0.199319)(0.326, 0.199969) 
(0.327, 0.200619)(0.328, 0.201269)(0.329, 0.20192)(0.33, 
0.202571)(0.331, 0.203222)(0.332, 0.203873)(0.333, 0.204524) 
(0.334, 0.205176)(0.335, 0.205828)(0.336, 0.206479)(0.337, 
0.207132)(0.338, 0.207784)(0.339, 0.208436)(0.34, 0.209089) 
(0.341, 0.209742)(0.342, 0.210395)(0.343, 0.211048)(0.344, 
0.211701)(0.345, 0.212355)(0.346, 0.213009)(0.347, 0.213663) 
(0.348, 0.214317)(0.349, 0.214971)(0.35, 0.215626)(0.351, 
0.21628)(0.352, 0.216935)(0.353, 0.21759)(0.354, 0.218246) 
(0.355, 0.218901)(0.356, 0.219557)(0.357, 0.220213)(0.358, 
0.220869)(0.359, 0.221525)(0.36, 0.222181)(0.361, 0.222838) 
(0.362, 0.223494)(0.363, 0.224151)(0.364, 0.224809)(0.365, 
0.225466)(0.366, 0.226123)(0.367, 0.226781)(0.368, 0.227439) 
(0.369, 0.228097)(0.37, 0.228755)(0.371, 0.229414)(0.372, 
0.230072)(0.373, 0.230731)(0.374, 0.23139)(0.375, 0.23205) 
(0.376, 0.232709)(0.377, 0.233369)(0.378, 0.234028)(0.379, 
0.234688)(0.38, 0.235348)(0.381, 0.236009)(0.382, 0.236669) 
(0.383, 0.23733)(0.384, 0.237991)(0.385, 0.238652)(0.386, 
0.239314)(0.387, 0.239975)(0.388, 0.240637)(0.389, 0.241299) 
(0.39, 0.241961)(0.391, 0.242623)(0.392, 0.243286)(0.393, 
0.243948)(0.394, 0.244611)(0.395, 0.245274)(0.396, 0.245937) 
(0.397, 0.246601)(0.398, 0.247265)(0.399, 0.247928)(0.4, 
0.248592)(0.401, 0.249257)(0.402, 0.249921)(0.403, 0.250586) 
(0.404, 0.251251)(0.405, 0.251916)(0.406, 0.252581)(0.407, 
0.253246)(0.408, 0.253912)(0.409, 0.254578)(0.41, 0.255244) 
(0.411, 0.25591)(0.412, 0.256577)(0.413, 0.257243)(0.414, 
0.25791)(0.415, 0.258577)(0.416, 0.259244)(0.417, 0.259912) 
(0.418, 0.260579)(0.419, 0.261247)(0.42, 0.261915)(0.421, 
0.262583)(0.422, 0.263251)(0.423, 0.263919)(0.424, 0.264587) 
(0.425, 0.265255)(0.426, 0.265924)(0.427, 0.266593)(0.428, 
0.267262)(0.429, 0.267932)(0.43, 0.268601)(0.431, 0.269271) 
(0.432, 0.269941)(0.433, 0.270611)(0.434, 0.271282)(0.435, 
0.271953)(0.436, 0.272624)(0.437, 0.273295)(0.438, 0.273966) 
(0.439, 0.274638)(0.44, 0.27531)(0.441, 0.275982)(0.442, 
0.276654)(0.443, 0.277327)(0.444, 0.277999)(0.445, 0.278672) 
(0.446, 0.279346)(0.447, 0.280019)(0.448, 0.280693)(0.449, 
0.281367)(0.45, 0.282041)(0.451, 0.282715)(0.452, 0.28339) 
(0.453, 0.284065)(0.454, 0.28474)(0.455, 0.285415)(0.456, 
0.286091)(0.457, 0.286766)(0.458, 0.287442)(0.459, 0.288119) 
(0.46, 0.288795)(0.461, 0.289472)(0.462, 0.290149)(0.463, 
0.290826)(0.464, 0.291503)(0.465, 0.292181)(0.466, 0.292859) 
(0.467, 0.293537)(0.468, 0.294216)(0.469, 0.294894)(0.47, 
0.295573)(0.471, 0.296252)(0.472, 0.296931)(0.473, 0.297611) 
(0.474, 0.298291)(0.475, 0.298971)(0.476, 0.299651)(0.477, 
0.300332)(0.478, 0.301013)(0.479, 0.301694)(0.48, 0.302375) 
(0.481, 0.303056)(0.482, 0.303738)(0.483, 0.30442)(0.484, 
0.305102)(0.485, 0.305785)(0.486, 0.306468)(0.487, 0.307151) 
(0.488, 0.307834)(0.489, 0.308518)(0.49, 0.309201)(0.491, 
0.309885)(0.492, 0.31057)(0.493, 0.311254)(0.494, 0.311939) 
(0.495, 0.312624)(0.496, 0.313309)(0.497, 0.313995)(0.498, 
0.314681)(0.499, 0.315367)(0.5, 0.316053)(0.501, 0.316739) 
(0.502, 0.317426)(0.503, 0.318113)(0.504, 0.318801)(0.505, 
0.319488)(0.506, 0.320176)(0.507, 0.320864)(0.508, 0.321553) 
(0.509, 0.322241)(0.51, 0.32293)(0.511, 0.323619)(0.512, 
0.324309)(0.513, 0.324998)(0.514, 0.325688)(0.515, 0.326379) 
(0.516, 0.327069)(0.517, 0.32776)(0.518, 0.328451)(0.519, 
0.329142)(0.52, 0.329834)(0.521, 0.330526)(0.522, 0.331218) 
(0.523, 0.33191)(0.524, 0.332603)(0.525, 0.333296)(0.526, 
0.333989)(0.527, 0.334682)(0.528, 0.335376)(0.529, 0.33607) 
(0.53, 0.336764)(0.531, 0.337459)(0.532, 0.338154)(0.533, 
0.338849)(0.534, 0.339544)(0.535, 0.34024)(0.536, 0.340936) 
(0.537, 0.341632)(0.538, 0.342328)(0.539, 0.343025)(0.54, 
0.343722)(0.541, 0.34442)(0.542, 0.345117)(0.543, 0.345815) 
(0.544, 0.346514)(0.545, 0.347212)(0.546, 0.347911)(0.547, 
0.34861)(0.548, 0.34931)(0.549, 0.350009)(0.55, 0.350709) 
(0.551, 0.35141)(0.552, 0.35211)(0.553, 0.352811)(0.554, 
0.353512)(0.555, 0.354214)(0.556, 0.354915)(0.557, 0.355617) 
(0.558, 0.35632)(0.559, 0.357022)(0.56, 0.357725)(0.561, 
0.358428)(0.562, 0.359132)(0.563, 0.359835)(0.564, 0.360539) 
(0.565, 0.361244)(0.566, 0.361948)(0.567, 0.362653)(0.568, 
0.363359)(0.569, 0.364064)(0.57, 0.36477)(0.571, 0.365476) 
(0.572, 0.366183)(0.573, 0.366889)(0.574, 0.367596)(0.575, 
0.368304)(0.576, 0.369011)(0.577, 0.369719)(0.578, 0.370428) 
(0.579, 0.371136)(0.58, 0.371845)(0.581, 0.372554)(0.582, 
0.373264)(0.583, 0.373974)(0.584, 0.374684)(0.585, 0.375394) 
(0.586, 0.376105)(0.587, 0.376816)(0.588, 0.377528)(0.589, 
0.378239)(0.59, 0.378951)(0.591, 0.379664)(0.592, 0.380377) 
(0.593, 0.38109)(0.594, 0.381803)(0.595, 0.382517)(0.596, 
0.38323)(0.597, 0.383945)(0.598, 0.384659)(0.599, 0.385374) 
(0.6, 0.38609)(0.601, 0.386805)(0.602, 0.387521)(0.603, 
0.388237)(0.604, 0.388954)(0.605, 0.389671)(0.606, 0.390388) 
(0.607, 0.391106)(0.608, 0.391824)(0.609, 0.392542)(0.61, 
0.393261)(0.611, 0.39398)(0.612, 0.394699)(0.613, 0.395419) 
(0.614, 0.396139)(0.615, 0.396859)(0.616, 0.39758)(0.617, 
0.398301)(0.618, 0.399022)(0.619, 0.399744)(0.62, 0.400466) 
(0.621, 0.401189)(0.622, 0.401911)(0.623, 0.402635)(0.624, 
0.403358)(0.625, 0.404082)(0.626, 0.404806)(0.627, 0.40553) 
(0.628, 0.406255)(0.629, 0.406981)(0.63, 0.407706)(0.631, 
0.408432)(0.632, 0.409158)(0.633, 0.409885)(0.634, 0.410612) 
(0.635, 0.411339)(0.636, 0.412067)(0.637, 0.412795)(0.638, 
0.413524)(0.639, 0.414253)(0.64, 0.414982)(0.641, 0.415711) 
(0.642, 0.416441)(0.643, 0.417172)(0.644, 0.417902)(0.645, 
0.418634)(0.646, 0.419365)(0.647, 0.420097)(0.648, 0.420829) 
(0.649, 0.421562)(0.65, 0.422295)(0.651, 0.423028)(0.652, 
0.423762)(0.653, 0.424496)(0.654, 0.425231)(0.655, 0.425965) 
(0.656, 0.426701)(0.657, 0.427436)(0.658, 0.428173)(0.659, 
0.428909)(0.66, 0.429646)(0.661, 0.430383)(0.662, 0.431121) 
(0.663, 0.431859)(0.664, 0.432597)(0.665, 0.433336)(0.666, 
0.434075)(0.667, 0.434815)(0.668, 0.435555)(0.669, 0.436296) 
(0.67, 0.437036)(0.671, 0.437778)(0.672, 0.438519)(0.673, 
0.439262)(0.674, 0.440004)(0.675, 0.440747)(0.676, 0.44149) 
(0.677, 0.442234)(0.678, 0.442978)(0.679, 0.443723)(0.68, 
0.444468)(0.681, 0.445213)(0.682, 0.445959)(0.683, 0.446706) 
(0.684, 0.447452)(0.685, 0.448199)(0.686, 0.448947)(0.687, 
0.449695)(0.688, 0.450443)(0.689, 0.451192)(0.69, 0.451942) 
(0.691, 0.452691)(0.692, 0.453442)(0.693, 0.454192)(0.694, 
0.454943)(0.695, 0.455695)(0.696, 0.456447)(0.697, 0.457199) 
(0.698, 0.457952)(0.699, 0.458706)(0.7, 0.459459)
};
 \end{axis}
\begin{axis}[anchor=north west ,xshift=0.5cm, yshift=5.5cm, width=0.25\textwidth, xmin=0, xmax=10, xlabel=$t\omega_\text{f}$, ylabel=$(\Theta(t)-T_0)/\Delta T$,style={font=\scriptsize}, ylabel near ticks, yticklabel pos=right, xlabel near ticks]
\addplot[color=blue] coordinates {
(0.001, 0.772669)(0.051, 0.71817)(0.101, 0.698279)(0.151,
0.684626)(0.201, 0.674044)(0.251, 0.665343)(0.301, 0.657931)
(0.351, 0.651463)(0.401, 0.645722)(0.451, 0.640559)(0.501,
0.635867)(0.551, 0.631568)(0.601, 0.627601)(0.651, 0.623919)
(0.701, 0.620486)(0.751, 0.61727)(0.801, 0.614247)(0.851,
0.611395)(0.901, 0.608697)(0.951, 0.606137)(1.001, 0.603704)
(1.051, 0.601385)(1.101, 0.599172)(1.151, 0.597054)(1.201,
0.595026)(1.251, 0.59308)(1.301, 0.59121)(1.351, 0.589411)
(1.401, 0.587678)(1.451, 0.586007)(1.501, 0.584393)(1.551,
0.582834)(1.601, 0.581326)(1.651, 0.579866)(1.701, 0.578452)
(1.751, 0.57708)(1.801, 0.575749)(1.851, 0.574457)(1.901,
0.573201)(1.951, 0.571979)(2.001, 0.570791)(2.051, 0.569634)
(2.101, 0.568508)(2.151, 0.567409)(2.201, 0.566339)(2.251,
0.565294)(2.301, 0.564274)(2.351, 0.563277)(2.401, 0.562304)
(2.451, 0.561352)(2.501, 0.560421)(2.551, 0.559511)(2.601,
0.55862)(2.651, 0.557749)(2.701, 0.556895)(2.751, 0.55606)
(2.801, 0.555241)(2.851, 0.55444)(2.901, 0.553655)(2.951,
0.552887)(3.001, 0.552134)(3.051, 0.551393)(3.101, 0.550666)
(3.151, 0.549952)(3.201, 0.549251)(3.251, 0.548562)(3.301,
0.547885)(3.351, 0.547219)(3.401, 0.546565)(3.451, 0.545922)
(3.501, 0.545289)(3.551, 0.544667)(3.601, 0.544054)(3.651,
0.543452)(3.701, 0.542858)(3.751, 0.542274)(3.801, 0.541699)
(3.851, 0.541132)(3.901, 0.540574)(3.951, 0.540024)(4.001,
0.539483)(4.051, 0.53895)(4.101, 0.538425)(4.151, 0.537907)
(4.201, 0.537397)(4.251, 0.536894)(4.301, 0.536398)(4.351,
0.535909)(4.401, 0.535426)(4.451, 0.53495)(4.501, 0.534481)
(4.551, 0.534017)(4.601, 0.53356)(4.651, 0.533109)(4.701,
0.532663)(4.751, 0.532224)(4.801, 0.531789)(4.851, 0.53136)
(4.901, 0.530937)(4.951, 0.530519)(5.001, 0.530105)(5.051,
0.529697)(5.101, 0.529293)(5.151, 0.528894)(5.201, 0.5285)
(5.251, 0.528111)(5.301, 0.527726)(5.351, 0.527345)(5.401,
0.526969)(5.451, 0.526597)(5.501, 0.52623)(5.551, 0.525866)
(5.601, 0.525507)(5.651, 0.525151)(5.701, 0.524799)(5.751,
0.524451)(5.801, 0.524107)(5.851, 0.523766)(5.901, 0.52343)
(5.951, 0.523096)(6.001, 0.522766)(6.051, 0.522439)(6.101,
0.522116)(6.151, 0.521795)(6.201, 0.521478)(6.251, 0.521164)
(6.301, 0.520854)(6.351, 0.520546)(6.401, 0.520241)(6.451,
0.51994)(6.501, 0.519641)(6.551, 0.519345)(6.601, 0.519052)
(6.651, 0.518762)(6.701, 0.518474)(6.751, 0.51819)(6.801,
0.517907)(6.851, 0.517628)(6.901, 0.517351)(6.951, 0.517076)
(7.001, 0.516805)(7.051, 0.516535)(7.101, 0.516267)(7.151,
0.516002)(7.201, 0.51574)(7.251, 0.515479)(7.301, 0.515221)
(7.351, 0.514965)(7.401, 0.514712)(7.451, 0.51446)(7.501,
0.514211)(7.551, 0.513964)(7.601, 0.513718)(7.651, 0.513475)
(7.701, 0.513234)(7.751, 0.512995)(7.801, 0.512758)(7.851,
0.512523)(7.901, 0.51229)(7.951, 0.512058)(8.001, 0.511829)
(8.051, 0.511601)(8.101, 0.511375)(8.151, 0.511151)(8.201,
0.510928)(8.251, 0.510707)(8.301, 0.510488)(8.351, 0.510271)
(8.401, 0.510055)(8.451, 0.509841)(8.501, 0.509628)(8.551,
0.509418)(8.601, 0.509208)(8.651, 0.509001)(8.701, 0.508795)
(8.751, 0.50859)(8.801, 0.508387)(8.851, 0.508185)(8.901,
0.507985)(8.951, 0.507786)(9.001, 0.507589)(9.051, 0.507393)
(9.101, 0.507199)(9.151, 0.507005)(9.201, 0.506814)(9.251,
0.506623)(9.301, 0.506434)(9.351, 0.506246)(9.401, 0.506059)
(9.451, 0.505874)(9.501, 0.50569)(9.551, 0.505508)(9.601,
0.505326)(9.651, 0.505146)(9.701, 0.504967)(9.751, 0.504789)
(9.801, 0.504612)(9.851, 0.504437)(9.901, 0.504262)(9.951,
0.504089)(10.001, 0.503917)
};
\end{axis}
\end{tikzpicture}
\caption{Parametric plot of the integrated response function $\mu(t)$
  against the time-dependent diffusivity
  $D(t)$ for a hot Brownian particle. Dashed lines indicate the
  asymptotic slopes $1/T^\V$ and $1/\caT(0)$ for short and long times, respectively. 
\emph{Inset:} the fluctuation dissipation ratio $\Theta(t)$ from
Eq.~(\ref{eq:caX}), i.e., the inverse slope of the curve in
the main plot, illustrating the crossover of the effective temperature
(normalized to the temperature difference $\Delta T$ between the particle surface
  and the ambient temperature $T_0$).}
\label{fig:parametric}
\end{figure}
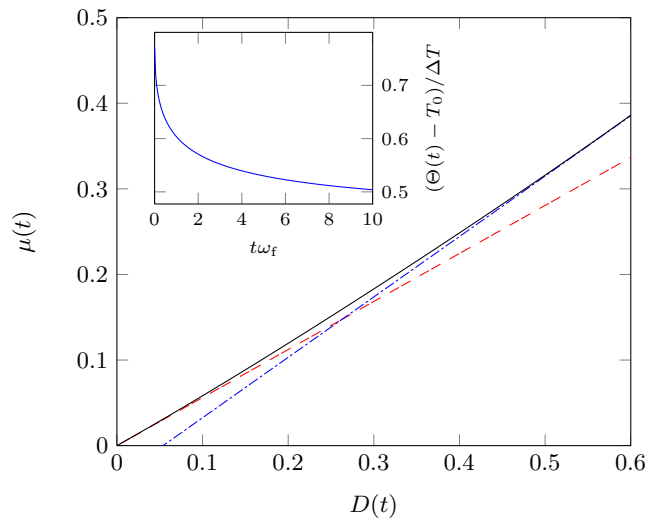

\emph{Fluctuation-dissipation relations}---The violation of the classical FDT of the second kind in
Eqs.~(\ref{noise}), (\ref{Tdef}) entails a corresponding violation of
the classical FDT of the first kind. To elucidate this point, we now focus 
on so-called hot Brownian motion \cite{Rings.2010}, the case where the Brownian particle itself
acts as the heat source and generates a co-moving radial temperature field $T(r)=T_0+\Delta T R/r$ in the fluid. 
Besides its practical relevance \cite{Selmke.2013},
this highly symmetric system may serve as a prototypical example to understand
the main implications of a frequency-dependent noise temperature
without the additional complications of the tensorial structure
and position dependence of the noise temperature. For the hot Brownian
particle Eqs.~\eqref{GLE}, \eqref{noise} imply
\begin{equation}\label{Cv}
  C_\V(\omega) = 2\kB \caT(\omega) \Real  R_\V(\omega)\,,
\end{equation}
where $C_\V(\omega)=\mean{V_i(\omega) V_i(-\omega)}$  is  the spectral
density and $R_\V(\omega)=(\zeta^\+(\omega)-i\omega m)^{-1}$ the
response function of the particle velocity
\cite{Falasco.2014}. The time-dependencies of the linear response and of the
spontaneous fluctuations of the Brownian particle obviously cannot be
identified, as in equilibrium. Hence, the classical FDT is broken and
Onsager's regression hypothesis is not applicable. As in quantum
mechanics, a single value of the noise temperature is generally not
sufficient to fully characterize the stochastic excitations provided
by the heat reservoir \cite{Ford.1996}. Instead, the energy carried by a degree of
freedom is given as a sum of filtered mode contributions, with the
filter bandwidth selected by its own response function. The fluctuation-dissipation ratio  
\begin{equation}\label{eq:caX}
\Theta(t)= D(t)/\mu(t)
\end{equation}
of the positional fluctuations and the response (to a small constant
force) no longer reduces to a universal constant $T_0$, as in an
isothermal bath.  Instead, for the hot particle, the time-dependent
diffusion coefficient $D(t)= \int_0^t C_\V(t') \text{d}t'$ and the
integrated response function $\mu(t)=\int_0^t R_\V(t') \text{d}t'$ only
become proportional to each other in asymptotic limits
(Fig.~\ref{fig:parametric}). More
precisely, it follows from Eq.~\eqref{Cv} that $\Theta(t)$ interpolates
between the two effective temperatures \cite{Falasco.2014}
\begin{equation}
T^\V= \frac{M}{\kB} \mean{V_i^2}= \frac{M}{\pi} \int \diff \omega
\caT(\omega) \Real  R_\V(\omega)
\end{equation}
at short times, and 
\begin{equation}
T^\X = \caT(0)< T^\V
\end{equation}
at long times, in agreement with what was found in numerical simulations \cite{Barrat.2011,
  Chakraborty.2011}. The generalization of Eqs.~\eqref{eq:einstein} for a hot particle is thus
\begin{equation}\label{eq:epipartition}
  \begin{split}
    \underset{t\to\infty}\lim \avg{[\vec X(t)- \vec X(0)]^2}/6t &
    = 3\kB T^\X\!/\zeta\\  
     \underset{t\to 0}\lim \avg{\vec V(t) \cdot\vec V(0)} & = 3\kB
     T^\V\!/M \,.
  \end{split}
\end{equation}

Along the same lines, one can discuss a hot Brownian particle in harmonic
confinement, a scenario of considerable practical relevance to metal
nanoparticles controlled by optical tweezers \cite{Juan.2011}.
Choosing the positional degree of freedom $X_i$ as our observable and
introducing the external confinement force $-m\omega_0^2 X_i(t)$ into
Eq.~\eqref{GLE}, one obtains the equivalent of the velocity equation
\eqref{Cv} for the particle position,
\begin{equation}\label{Cx}
\omega C_\X(\omega) = 2\kB \caT(\omega) \Imag  R_\X(\omega)\,.
\end{equation}
Here, $C_\X(\omega)=\mean{X_i(\omega)X_i(-\omega)}$ is now the spectral density and
$R_\X(\omega)=(m(\omega_0^2-\omega^2)-i\omega \zeta^\+(\omega))^{-1}$
the response function of the position coordinate.  Since the response functions $R_\V(\omega)$ and
$R_\X(\omega)$ for the velocity and the position filter the overall
temperature spectrum $\caT(\omega)$ differently, energy
equipartition between the (harmonic) position and momentum degrees of
freedom inevitably breaks down. Velocity and position fluctuations 
of a hot Brownian harmonic oscillator thus
thermalize to different effective temperatures 
\begin{equation}
  m\omega_0^2 \mean{X_i^2}= \kB T^\X \, \qquad  M\mean{V_i^2}= \kB T^\V \;.
\end{equation}
We emphasize that these depend on the confinement strength for
sufficiently strong confinement/weak solvent coupling,
so that they generally cannot be identified with the corresponding
temperatures denoted by the same symbols in Eq.~(\ref{eq:epipartition}) \cite{Falasco.2014}.

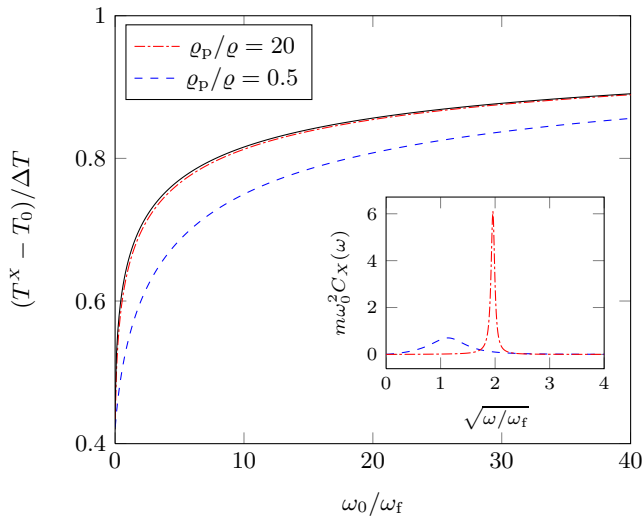
\begin{figure}[tb]
\begin{tikzpicture}
\begin{axis}[xmin=0,xmax=40,ymin=0.4 ,ymax=1.0,legend style={at={(0.02,0.98)}, anchor= north west}, xlabel=$\omega_0/\omega_\text{f}$,
ylabel={$(T^\X-T_0)/\Delta T$}] 
\addplot[color=red, dash pattern=on 4pt off 1pt on 1pt off 1pt] coordinates{
(0.01,0.434545)(0.11,0.503545)(0.21,0.538118)(0.31,0.561956)(0.41,0.58029)(0.51,0.595229)(0.61,0.607846)(0.71,0.618768)(0.81,0.628396)(0.91,0.637002)(1.01,0.644779)(1.11,0.651869)(1.21,0.658381)(1.31,0.664399)(1.41,0.669991)(1.51,0.675211)(1.61,0.680104)(1.71,0.684706)(1.81,0.689048)(1.91,0.693157)(2.01,0.697056)(2.11,0.700763)(2.21,0.704295)(2.31,0.707668)(2.41,0.710894)(2.51,0.713985)(2.61,0.716951)(2.71,0.7198)(2.81,0.722542)(2.91,0.725183)(3.01,0.72773)(3.11,0.730189)(3.21,0.732566)(3.31,0.734865)(3.41,0.737091)(3.51,0.739247)(3.61,0.741339)(3.71,0.743369)(3.81,0.745341)(3.91,0.747257)(4.01,0.749121)(4.11,0.750935)(4.21,0.752701)(4.31,0.754421)(4.41,0.756098)(4.51,0.757734)(4.61,0.75933)(4.71,0.760888)(4.81,0.76241)(4.91,0.763897)(5.01,0.765351)(5.11,0.766773)(5.21,0.768164)(5.31,0.769525)(5.41,0.770858)(5.51,0.772163)(5.61,0.773442)(5.71,0.774696)(5.81,0.775924)(5.91,0.777129)(6.01,0.778311)(6.11,0.779471)(6.21,0.78061)(6.31,0.781728)(6.41,0.782825)(6.51,0.783904)(6.61,0.784963)(6.71,0.786004)(6.81,0.787028)(6.91,0.788034)(7.01,0.789024)(7.11,0.789997)(7.21,0.790955)(7.31,0.791898)(7.41,0.792825)(7.51,0.793739)(7.61,0.794638)(7.71,0.795523)(7.81,0.796396)(7.91,0.797255)(8.01,0.798102)(8.11,0.798937)(8.21,0.799759)(8.31,0.80057)(8.41,0.80137)(8.51,0.802158)(8.61,0.802936)(8.71,0.803703)(8.81,0.80446)(8.91,0.805207)(9.01,0.805943)(9.11,0.806671)(9.21,0.807389)(9.31,0.808097)(9.41,0.808797)(9.51,0.809488)(9.61,0.810171)(9.71,0.810845)(9.81,0.811511)(9.91,0.812168)(10.01,0.812818)(10.11,0.813461)(10.21,0.814095)(10.31,0.814723)(10.41,0.815343)(10.51,0.815956)(10.61,0.816562)(10.71,0.817161)(10.81,0.817754)(10.91,0.81834)(11.01,0.818919)(11.11,0.819493)(11.21,0.82006)(11.31,0.820621)(11.41,0.821176)(11.51,0.821725)(11.61,0.822269)(11.71,0.822807)(11.81,0.823339)(11.91,0.823866)(12.01,0.824388)(12.11,0.824904)(12.21,0.825416)(12.31,0.825922)(12.41,0.826423)(12.51,0.82692)(12.61,0.827411)(12.71,0.827898)(12.81,0.82838)(12.91,0.828858)(13.01,0.829331)(13.11,0.8298)(13.21,0.830264)(13.31,0.830724)(13.41,0.83118)(13.51,0.831632)(13.61,0.83208)(13.71,0.832524)(13.81,0.832963)(13.91,0.833399)(14.01,0.833831)(14.11,0.83426)(14.21,0.834684)(14.31,0.835105)(14.41,0.835522)(14.51,0.835936)(14.61,0.836347)(14.71,0.836753)(14.81,0.837157)(14.91,0.837557)(15.01,0.837954)(15.11,0.838347)(15.21,0.838738)(15.31,0.839125)(15.41,0.839509)(15.51,0.83989)(15.61,0.840268)(15.71,0.840643)(15.81,0.841015)(15.91,0.841385)(16.01,0.841751)(16.11,0.842114)(16.21,0.842475)(16.31,0.842833)(16.41,0.843188)(16.51,0.843541)(16.61,0.843891)(16.71,0.844238)(16.81,0.844583)(16.91,0.844925)(17.01,0.845265)(17.11,0.845602)(17.21,0.845937)(17.31,0.846269)(17.41,0.846599)(17.51,0.846927)(17.61,0.847252)(17.71,0.847575)(17.81,0.847896)(17.91,0.848214)(18.01,0.84853)(18.11,0.848845)(18.21,0.849156)(18.31,0.849466)(18.41,0.849774)(18.51,0.85008)(18.61,0.850383)(18.71,0.850685)(18.81,0.850984)(18.91,0.851281)(19.01,0.851577)(19.11,0.851871)(19.21,0.852162)(19.31,0.852452)(19.41,0.85274)(19.51,0.853026)(19.61,0.85331)(19.71,0.853592)(19.81,0.853873)(19.91,0.854152)(20.01,0.854429)(20.11,0.854704)(20.21,0.854978)(20.31,0.855249)(20.41,0.85552)(20.51,0.855788)(20.61,0.856055)(20.71,0.85632)(20.81,0.856584)(20.91,0.856846)(21.01,0.857106)(21.11,0.857365)(21.21,0.857622)(21.31,0.857878)(21.41,0.858133)(21.51,0.858385)(21.61,0.858637)(21.71,0.858886)(21.81,0.859135)(21.91,0.859382)(22.01,0.859627)(22.11,0.859871)(22.21,0.860114)(22.31,0.860355)(22.41,0.860595)(22.51,0.860834)(22.61,0.861071)(22.71,0.861307)(22.81,0.861542)(22.91,0.861775)(23.01,0.862007)(23.11,0.862238)(23.21,0.862467)(23.31,0.862696)(23.41,0.862923)(23.51,0.863148)(23.61,0.863373)(23.71,0.863596)(23.81,0.863818)(23.91,0.864039)(24.01,0.864259)(24.11,0.864477)(24.21,0.864695)(24.31,0.864911)(24.41,0.865126)(24.51,0.86534)(24.61,0.865553)(24.71,0.865765)(24.81,0.865976)(24.91,0.866185)(25.01,0.866394)(25.11,0.866602)(25.21,0.866808)(25.31,0.867013)(25.41,0.867218)(25.51,0.867421)(25.61,0.867623)(25.71,0.867825)(25.81,0.868025)(25.91,0.868224)(26.01,0.868423)(26.11,0.86862)(26.21,0.868816)(26.31,0.869012)(26.41,0.869206)(26.51,0.8694)(26.61,0.869592)(26.71,0.869784)(26.81,0.869975)(26.91,0.870165)(27.01,0.870354)(27.11,0.870542)(27.21,0.870729)(27.31,0.870915)(27.41,0.8711)(27.51,0.871285)(27.61,0.871469)(27.71,0.871651)(27.81,0.871833)(27.91,0.872015)(28.01,0.872195)(28.11,0.872374)(28.21,0.872553)(28.31,0.872731)(28.41,0.872908)(28.51,0.873084)(28.61,0.873259)(28.71,0.873434)(28.81,0.873608)(28.91,0.873781)(29.01,0.873953)(29.11,0.874125)(29.21,0.874295)(29.31,0.874465)(29.41,0.874635)(29.51,0.874803)(29.61,0.874971)(29.71,0.875138)(29.81,0.875304)(29.91,0.87547)(30.01,0.875635)(30.11,0.875799)(30.21,0.875962)(30.31,0.876125)(30.41,0.876287)(30.51,0.876449)(30.61,0.876609)(30.71,0.876769)(30.81,0.876929)(30.91,0.877088)(31.01,0.877246)(31.11,0.877403)(31.21,0.87756)(31.31,0.877716)(31.41,0.877871)(31.51,0.878026)(31.61,0.87818)(31.71,0.878333)(31.81,0.878486)(31.91,0.878639)(32.01,0.87879)(32.11,0.878941)(32.21,0.879092)(32.31,0.879241)(32.41,0.879391)(32.51,0.879539)(32.61,0.879687)(32.71,0.879835)(32.81,0.879981)(32.91,0.880128)(33.01,0.880273)(33.11,0.880419)(33.21,0.880563)(33.31,0.880707)(33.41,0.88085)(33.51,0.880993)(33.61,0.881136)(33.71,0.881277)(33.81,0.881418)(33.91,0.881559)(34.01,0.881699)(34.11,0.881839)(34.21,0.881978)(34.31,0.882116)(34.41,0.882254)(34.51,0.882392)(34.61,0.882529)(34.71,0.882665)(34.81,0.882801)(34.91,0.882936)(35.01,0.883071)(35.11,0.883206)(35.21,0.88334)(35.31,0.883473)(35.41,0.883606)(35.51,0.883738)(35.61,0.88387)(35.71,0.884002)(35.81,0.884133)(35.91,0.884263)(36.01,0.884393)(36.11,0.884523)(36.21,0.884652)(36.31,0.88478)(36.41,0.884909)(36.51,0.885036)(36.61,0.885164)(36.71,0.88529)(36.81,0.885417)(36.91,0.885543)(37.01,0.885668)(37.11,0.885793)(37.21,0.885917)(37.31,0.886042)(37.41,0.886165)(37.51,0.886288)(37.61,0.886411)(37.71,0.886534)(37.81,0.886656)(37.91,0.886777)(38.01,0.886898)(38.11,0.887019)(38.21,0.887139)(38.31,0.887259)(38.41,0.887379)(38.51,0.887498)(38.61,0.887616)(38.71,0.887734)(38.81,0.887852)(38.91,0.88797)(39.01,0.888087)(39.11,0.888203)(39.21,0.88832)(39.31,0.888436)(39.41,0.888551)(39.51,0.888666)(39.61,0.888781)(39.71,0.888895)(39.81,0.889009)(39.91,0.889123)(40.01,0.889236)(40.11,0.889349)(40.21,0.889461)(40.31,0.889573)(40.41,0.889685)(40.51,0.889796)(40.61,0.889907)(40.71,0.890018)(40.81,0.890128)(40.91,0.890238)(41.01,0.890348)(41.11,0.890457)(41.21,0.890566)(41.31,0.890674)(41.41,0.890783)(41.51,0.89089)(41.61,0.890998)(41.71,0.891105)(41.81,0.891212)(41.91,0.891318)(42.01,0.891425)(42.11,0.89153)(42.21,0.891636)(42.31,0.891741)(42.41,0.891846)(42.51,0.89195)(42.61,0.892054)(42.71,0.892158)(42.81,0.892262)(42.91,0.892365)(43.01,0.892468)(43.11,0.89257)(43.21,0.892673)(43.31,0.892775)(43.41,0.892876)(43.51,0.892978)(43.61,0.893079)(43.71,0.893179)(43.81,0.89328)(43.91,0.89338)(44.01,0.89348)(44.11,0.893579)(44.21,0.893678)(44.31,0.893777)(44.41,0.893876)(44.51,0.893974)(44.61,0.894072)(44.71,0.89417)(44.81,0.894268)(44.91,0.894365)(45.01,0.894462)(45.11,0.894558)(45.21,0.894654)(45.31,0.89475)(45.41,0.894846)(45.51,0.894942)(45.61,0.895037)(45.71,0.895132)(45.81,0.895226)(45.91,0.895321)(46.01,0.895415)(46.11,0.895509)(46.21,0.895602)(46.31,0.895696)(46.41,0.895789)(46.51,0.895881)(46.61,0.895974)(46.71,0.896066)(46.81,0.896158)(46.91,0.89625)(47.01,0.896341)(47.11,0.896432)(47.21,0.896523)(47.31,0.896614)(47.41,0.896704)(47.51,0.896794)(47.61,0.896884)(47.71,0.896974)(47.81,0.897063)(47.91,0.897153)(48.01,0.897241)(48.11,0.89733)(48.21,0.897419)(48.31,0.897507)(48.41,0.897595)(48.51,0.897682)(48.61,0.89777)(48.71,0.897857)(48.81,0.897944)(48.91,0.898031)(49.01,0.898117)(49.11,0.898204)(49.21,0.89829)(49.31,0.898375)(49.41,0.898461)(49.51,0.898546)(49.61,0.898631)(49.71,0.898716)(49.81,0.898801)(49.91,0.898885)
};\addplot[color=blue,dashed] coordinates{
(0.01,0.423009)(0.11,0.444179)(0.21,0.46118)(0.31,0.475696)(0.41,0.488453)(0.51,0.499869)(0.61,0.510217)(0.71,0.519689)(0.81,0.528427)(0.91,0.536536)(1.01,0.544103)(1.11,0.551195)(1.21,0.557867)(1.31,0.564166)(1.41,0.570129)(1.51,0.575789)(1.61,0.581176)(1.71,0.586312)(1.81,0.59122)(1.91,0.595916)(2.01,0.600419)(2.11,0.604742)(2.21,0.608899)(2.31,0.6129)(2.41,0.616757)(2.51,0.620478)(2.61,0.624073)(2.71,0.627548)(2.81,0.630911)(2.91,0.634169)(3.01,0.637327)(3.11,0.640391)(3.21,0.643366)(3.31,0.646255)(3.41,0.649065)(3.51,0.651798)(3.61,0.654458)(3.71,0.657049)(3.81,0.659574)(3.91,0.662036)(4.01,0.664437)(4.11,0.666781)(4.21,0.669069)(4.31,0.671304)(4.41,0.673489)(4.51,0.675624)(4.61,0.677713)(4.71,0.679757)(4.81,0.681757)(4.91,0.683715)(5.01,0.685633)(5.11,0.687513)(5.21,0.689354)(5.31,0.69116)(5.41,0.692931)(5.51,0.694668)(5.61,0.696373)(5.71,0.698046)(5.81,0.699688)(5.91,0.701301)(6.01,0.702886)(6.11,0.704442)(6.21,0.705972)(6.31,0.707476)(6.41,0.708954)(6.51,0.710408)(6.61,0.711838)(6.71,0.713245)(6.81,0.714629)(6.91,0.715991)(7.01,0.717333)(7.11,0.718653)(7.21,0.719953)(7.31,0.721234)(7.41,0.722495)(7.51,0.723738)(7.61,0.724962)(7.71,0.726169)(7.81,0.727359)(7.91,0.728532)(8.01,0.729689)(8.11,0.730829)(8.21,0.731954)(8.31,0.733064)(8.41,0.734158)(8.51,0.735238)(8.61,0.736304)(8.71,0.737356)(8.81,0.738395)(8.91,0.73942)(9.01,0.740432)(9.11,0.741432)(9.21,0.742419)(9.31,0.743393)(9.41,0.744356)(9.51,0.745308)(9.61,0.746248)(9.71,0.747176)(9.81,0.748094)(9.91,0.749001)(10.01,0.749898)(10.11,0.750784)(10.21,0.75166)(10.31,0.752526)(10.41,0.753383)(10.51,0.75423)(10.61,0.755067)(10.71,0.755896)(10.81,0.756715)(10.91,0.757526)(11.01,0.758328)(11.11,0.759122)(11.21,0.759907)(11.31,0.760684)(11.41,0.761453)(11.51,0.762214)(11.61,0.762967)(11.71,0.763713)(11.81,0.764451)(11.91,0.765182)(12.01,0.765905)(12.11,0.766622)(12.21,0.767331)(12.31,0.768034)(12.41,0.76873)(12.51,0.769419)(12.61,0.770101)(12.71,0.770778)(12.81,0.771447)(12.91,0.772111)(13.01,0.772768)(13.11,0.77342)(13.21,0.774065)(13.31,0.774705)(13.41,0.775339)(13.51,0.775967)(13.61,0.77659)(13.71,0.777207)(13.81,0.777818)(13.91,0.778425)(14.01,0.779026)(14.11,0.779622)(14.21,0.780213)(14.31,0.780798)(14.41,0.781379)(14.51,0.781955)(14.61,0.782526)(14.71,0.783093)(14.81,0.783655)(14.91,0.784212)(15.01,0.784764)(15.11,0.785312)(15.21,0.785856)(15.31,0.786395)(15.41,0.78693)(15.51,0.787461)(15.61,0.787988)(15.71,0.78851)(15.81,0.789029)(15.91,0.789543)(16.01,0.790054)(16.11,0.79056)(16.21,0.791063)(16.31,0.791562)(16.41,0.792057)(16.51,0.792548)(16.61,0.793036)(16.71,0.79352)(16.81,0.794)(16.91,0.794477)(17.01,0.794951)(17.11,0.795421)(17.21,0.795888)(17.31,0.796351)(17.41,0.796811)(17.51,0.797268)(17.61,0.797721)(17.71,0.798171)(17.81,0.798619)(17.91,0.799063)(18.01,0.799504)(18.11,0.799941)(18.21,0.800376)(18.31,0.800808)(18.41,0.801237)(18.51,0.801663)(18.61,0.802086)(18.71,0.802507)(18.81,0.802924)(18.91,0.803339)(19.01,0.803751)(19.11,0.80416)(19.21,0.804567)(19.31,0.804971)(19.41,0.805372)(19.51,0.805771)(19.61,0.806167)(19.71,0.806561)(19.81,0.806952)(19.91,0.807341)(20.01,0.807727)(20.11,0.808111)(20.21,0.808492)(20.31,0.808871)(20.41,0.809248)(20.51,0.809622)(20.61,0.809994)(20.71,0.810364)(20.81,0.810731)(20.91,0.811097)(21.01,0.81146)(21.11,0.81182)(21.21,0.812179)(21.31,0.812536)(21.41,0.81289)(21.51,0.813242)(21.61,0.813593)(21.71,0.813941)(21.81,0.814287)(21.91,0.814631)(22.01,0.814973)(22.11,0.815313)(22.21,0.815652)(22.31,0.815988)(22.41,0.816322)(22.51,0.816655)(22.61,0.816985)(22.71,0.817314)(22.81,0.817641)(22.91,0.817966)(23.01,0.818289)(23.11,0.81861)(23.21,0.81893)(23.31,0.819248)(23.41,0.819564)(23.51,0.819878)(23.61,0.820191)(23.71,0.820502)(23.81,0.820811)(23.91,0.821119)(24.01,0.821425)(24.11,0.821729)(24.21,0.822032)(24.31,0.822333)(24.41,0.822633)(24.51,0.822931)(24.61,0.823227)(24.71,0.823522)(24.81,0.823816)(24.91,0.824107)(25.01,0.824398)(25.11,0.824686)(25.21,0.824974)(25.31,0.82526)(25.41,0.825544)(25.51,0.825827)(25.61,0.826109)(25.71,0.826389)(25.81,0.826667)(25.91,0.826945)(26.01,0.827221)(26.11,0.827495)(26.21,0.827768)(26.31,0.82804)(26.41,0.828311)(26.51,0.82858)(26.61,0.828848)(26.71,0.829114)(26.81,0.82938)(26.91,0.829644)(27.01,0.829906)(27.11,0.830168)(27.21,0.830428)(27.31,0.830687)(27.41,0.830945)(27.51,0.831201)(27.61,0.831456)(27.71,0.831711)(27.81,0.831963)(27.91,0.832215)(28.01,0.832466)(28.11,0.832715)(28.21,0.832963)(28.31,0.83321)(28.41,0.833456)(28.51,0.833701)(28.61,0.833945)(28.71,0.834188)(28.81,0.834429)(28.91,0.834669)(29.01,0.834909)(29.11,0.835147)(29.21,0.835384)(29.31,0.83562)(29.41,0.835855)(29.51,0.836089)(29.61,0.836322)(29.71,0.836554)(29.81,0.836785)(29.91,0.837015)(30.01,0.837244)(30.11,0.837472)(30.21,0.837699)(30.31,0.837925)(30.41,0.83815)(30.51,0.838374)(30.61,0.838597)(30.71,0.838819)(30.81,0.83904)(30.91,0.83926)(31.01,0.839479)(31.11,0.839698)(31.21,0.839915)(31.31,0.840131)(31.41,0.840347)(31.51,0.840562)(31.61,0.840775)(31.71,0.840988)(31.81,0.8412)(31.91,0.841411)(32.01,0.841622)(32.11,0.841831)(32.21,0.84204)(32.31,0.842247)(32.41,0.842454)(32.51,0.84266)(32.61,0.842865)(32.71,0.84307)(32.81,0.843273)(32.91,0.843476)(33.01,0.843678)(33.11,0.843879)(33.21,0.844079)(33.31,0.844278)(33.41,0.844477)(33.51,0.844675)(33.61,0.844872)(33.71,0.845068)(33.81,0.845264)(33.91,0.845459)(34.01,0.845653)(34.11,0.845846)(34.21,0.846039)(34.31,0.84623)(34.41,0.846421)(34.51,0.846612)(34.61,0.846801)(34.71,0.84699)(34.81,0.847178)(34.91,0.847366)(35.01,0.847552)(35.11,0.847738)(35.21,0.847924)(35.31,0.848108)(35.41,0.848292)(35.51,0.848475)(35.61,0.848658)(35.71,0.84884)(35.81,0.849021)(35.91,0.849201)(36.01,0.849381)(36.11,0.84956)(36.21,0.849739)(36.31,0.849917)(36.41,0.850094)(36.51,0.850271)(36.61,0.850447)(36.71,0.850622)(36.81,0.850796)(36.91,0.85097)(37.01,0.851144)(37.11,0.851317)(37.21,0.851489)(37.31,0.85166)(37.41,0.851831)(37.51,0.852001)(37.61,0.852171)(37.71,0.85234)(37.81,0.852509)(37.91,0.852677)(38.01,0.852844)(38.11,0.853011)(38.21,0.853177)(38.31,0.853342)(38.41,0.853507)(38.51,0.853672)(38.61,0.853836)(38.71,0.853999)(38.81,0.854162)(38.91,0.854324)(39.01,0.854485)(39.11,0.854646)(39.21,0.854807)(39.31,0.854967)(39.41,0.855126)(39.51,0.855285)(39.61,0.855443)(39.71,0.855601)(39.81,0.855758)(39.91,0.855915)(40.01,0.856071)(40.11,0.856227)(40.21,0.856382)(40.31,0.856537)(40.41,0.856691)(40.51,0.856844)(40.61,0.856997)(40.71,0.85715)(40.81,0.857302)(40.91,0.857454)(41.01,0.857605)(41.11,0.857755)(41.21,0.857906)(41.31,0.858055)(41.41,0.858204)(41.51,0.858353)(41.61,0.858501)(41.71,0.858649)(41.81,0.858796)(41.91,0.858943)(42.01,0.859089)(42.11,0.859235)(42.21,0.85938)(42.31,0.859525)(42.41,0.859669)(42.51,0.859813)(42.61,0.859957)(42.71,0.8601)(42.81,0.860243)(42.91,0.860385)(43.01,0.860526)(43.11,0.860668)(43.21,0.860808)(43.31,0.860949)(43.41,0.861089)(43.51,0.861228)(43.61,0.861367)(43.71,0.861506)(43.81,0.861644)(43.91,0.861782)(44.01,0.861919)(44.11,0.862056)(44.21,0.862193)(44.31,0.862329)(44.41,0.862465)(44.51,0.8626)(44.61,0.862735)(44.71,0.862869)(44.81,0.863003)(44.91,0.863137)(45.01,0.86327)(45.11,0.863403)(45.21,0.863536)(45.31,0.863668)(45.41,0.863799)(45.51,0.863931)(45.61,0.864061)(45.71,0.864192)(45.81,0.864322)(45.91,0.864452)(46.01,0.864581)(46.11,0.86471)(46.21,0.864839)(46.31,0.864967)(46.41,0.865095)(46.51,0.865222)(46.61,0.865349)(46.71,0.865476)(46.81,0.865602)(46.91,0.865728)(47.01,0.865854)(47.11,0.865979)(47.21,0.866104)(47.31,0.866229)(47.41,0.866353)(47.51,0.866477)(47.61,0.8666)(47.71,0.866723)(47.81,0.866846)(47.91,0.866969)(48.01,0.867091)(48.11,0.867212)(48.21,0.867334)(48.31,0.867455)(48.41,0.867576)(48.51,0.867696)(48.61,0.867816)(48.71,0.867936)(48.81,0.868055)(48.91,0.868174)(49.01,0.868293)(49.11,0.868411)(49.21,0.868529)(49.31,0.868647)(49.41,0.868764)(49.51,0.868881)(49.61,0.868998)(49.71,0.869115)(49.81,0.869231)(49.91,0.869347)
}; \addplot[color=black] coordinates{(0,0.42)(0.01,0.457333)(0.11,0.527029)(0.21,0.558047)(0.31,0.579508)(0.41,0.596136)(0.51,0.609775)(0.61,0.621358)(0.71,0.631433)(0.81,0.64035)(0.91,0.648346)(1.01,0.655593)(1.11,0.662217)(1.21,0.668315)(1.31,0.673963)(1.41,0.67922)(1.51,0.684135)(1.61,0.688749)(1.71,0.693095)(1.81,0.6972)(1.91,0.70109)(2.01,0.704785)(2.11,0.708301)(2.21,0.711656)(2.31,0.714861)(2.41,0.71793)(2.51,0.720872)(2.61,0.723697)(2.71,0.726414)(2.81,0.729029)(2.91,0.73155)(3.01,0.733983)(3.11,0.736333)(3.21,0.738605)(3.31,0.740805)(3.41,0.742935)(3.51,0.745)(3.61,0.747004)(3.71,0.74895)(3.81,0.750841)(3.91,0.752679)(4.01,0.754468)(4.11,0.756209)(4.21,0.757905)(4.31,0.759558)(4.41,0.76117)(4.51,0.762742)(4.61,0.764277)(4.71,0.765776)(4.81,0.767241)(4.91,0.768672)(5.01,0.770072)(5.11,0.771441)(5.21,0.772781)(5.31,0.774093)(5.41,0.775377)(5.51,0.776636)(5.61,0.777869)(5.71,0.779078)(5.81,0.780263)(5.91,0.781426)(6.01,0.782567)(6.11,0.783687)(6.21,0.784786)(6.31,0.785866)(6.41,0.786926)(6.51,0.787968)(6.61,0.788992)(6.71,0.789998)(6.81,0.790987)(6.91,0.79196)(7.01,0.792917)(7.11,0.793859)(7.21,0.794786)(7.31,0.795698)(7.41,0.796595)(7.51,0.797479)(7.61,0.79835)(7.71,0.799207)(7.81,0.800052)(7.91,0.800884)(8.01,0.801704)(8.11,0.802513)(8.21,0.80331)(8.31,0.804096)(8.41,0.80487)(8.51,0.805635)(8.61,0.806388)(8.71,0.807132)(8.81,0.807866)(8.91,0.80859)(9.01,0.809305)(9.11,0.81001)(9.21,0.810707)(9.31,0.811394)(9.41,0.812073)(9.51,0.812744)(9.61,0.813406)(9.71,0.81406)(9.81,0.814707)(9.91,0.815345)(10.01,0.815976)(10.11,0.8166)(10.21,0.817216)(10.31,0.817826)(10.41,0.818428)(10.51,0.819023)(10.61,0.819612)(10.71,0.820194)(10.81,0.82077)(10.91,0.82134)(11.01,0.821903)(11.11,0.82246)(11.21,0.823011)(11.31,0.823557)(11.41,0.824097)(11.51,0.824631)(11.61,0.825159)(11.71,0.825682)(11.81,0.8262)(11.91,0.826713)(12.01,0.82722)(12.11,0.827722)(12.21,0.82822)(12.31,0.828712)(12.41,0.8292)(12.51,0.829683)(12.61,0.830161)(12.71,0.830635)(12.81,0.831104)(12.91,0.831569)(13.01,0.83203)(13.11,0.832486)(13.21,0.832938)(13.31,0.833386)(13.41,0.83383)(13.51,0.83427)(13.61,0.834706)(13.71,0.835138)(13.81,0.835567)(13.91,0.835991)(14.01,0.836412)(14.11,0.836829)(14.21,0.837243)(14.31,0.837653)(14.41,0.83806)(14.51,0.838463)(14.61,0.838863)(14.71,0.839259)(14.81,0.839652)(14.91,0.840042)(15.01,0.840429)(15.11,0.840813)(15.21,0.841193)(15.31,0.841571)(15.41,0.841945)(15.51,0.842317)(15.61,0.842685)(15.71,0.843051)(15.81,0.843414)(15.91,0.843774)(16.01,0.844131)(16.11,0.844485)(16.21,0.844837)(16.31,0.845186)(16.41,0.845533)(16.51,0.845877)(16.61,0.846218)(16.71,0.846557)(16.81,0.846893)(16.91,0.847227)(17.01,0.847559)(17.11,0.847888)(17.21,0.848214)(17.31,0.848539)(17.41,0.848861)(17.51,0.84918)(17.61,0.849498)(17.71,0.849813)(17.81,0.850126)(17.91,0.850437)(18.01,0.850746)(18.11,0.851052)(18.21,0.851357)(18.31,0.851659)(18.41,0.851959)(18.51,0.852258)(18.61,0.852554)(18.71,0.852849)(18.81,0.853141)(18.91,0.853431)(19.01,0.85372)(19.11,0.854007)(19.21,0.854292)(19.31,0.854575)(19.41,0.854856)(19.51,0.855135)(19.61,0.855413)(19.71,0.855688)(19.81,0.855962)(19.91,0.856235)(20.01,0.856505)(20.11,0.856774)(20.21,0.857042)(20.31,0.857307)(20.41,0.857571)(20.51,0.857834)(20.61,0.858094)(20.71,0.858353)(20.81,0.858611)(20.91,0.858867)(21.01,0.859122)(21.11,0.859375)(21.21,0.859626)(21.31,0.859876)(21.41,0.860125)(21.51,0.860372)(21.61,0.860618)(21.71,0.860862)(21.81,0.861105)(21.91,0.861346)(22.01,0.861586)(22.11,0.861825)(22.21,0.862062)(22.31,0.862298)(22.41,0.862532)(22.51,0.862766)(22.61,0.862998)(22.71,0.863228)(22.81,0.863458)(22.91,0.863686)(23.01,0.863913)(23.11,0.864139)(23.21,0.864363)(23.31,0.864586)(23.41,0.864808)(23.51,0.865029)(23.61,0.865249)(23.71,0.865467)(23.81,0.865684)(23.91,0.8659)(24.01,0.866115)(24.11,0.866329)(24.21,0.866542)(24.31,0.866754)(24.41,0.866964)(24.51,0.867173)(24.61,0.867382)(24.71,0.867589)(24.81,0.867795)(24.91,0.868)(25.01,0.868204)(25.11,0.868407)(25.21,0.868609)(25.31,0.86881)(25.41,0.86901)(25.51,0.869209)(25.61,0.869407)(25.71,0.869604)(25.81,0.869801)(25.91,0.869996)(26.01,0.87019)(26.11,0.870383)(26.21,0.870575)(26.31,0.870766)(26.41,0.870957)(26.51,0.871146)(26.61,0.871335)(26.71,0.871522)(26.81,0.871709)(26.91,0.871895)(27.01,0.87208)(27.11,0.872264)(27.21,0.872447)(27.31,0.87263)(27.41,0.872811)(27.51,0.872992)(27.61,0.873172)(27.71,0.873351)(27.81,0.873529)(27.91,0.873706)(28.01,0.873883)(28.11,0.874058)(28.21,0.874233)(28.31,0.874408)(28.41,0.874581)(28.51,0.874753)(28.61,0.874925)(28.71,0.875096)(28.81,0.875267)(28.91,0.875436)(29.01,0.875605)(29.11,0.875773)(29.21,0.87594)(29.31,0.876107)(29.41,0.876272)(29.51,0.876438)(29.61,0.876602)(29.71,0.876766)(29.81,0.876928)(29.91,0.877091)(30.01,0.877252)(30.11,0.877413)(30.21,0.877573)(30.31,0.877733)(30.41,0.877892)(30.51,0.87805)(30.61,0.878207)(30.71,0.878364)(30.81,0.87852)(30.91,0.878676)(31.01,0.87883)(31.11,0.878985)(31.21,0.879138)(31.31,0.879291)(31.41,0.879443)(31.51,0.879595)(31.61,0.879746)(31.71,0.879897)(31.81,0.880046)(31.91,0.880196)(32.01,0.880344)(32.11,0.880492)(32.21,0.88064)(32.31,0.880787)(32.41,0.880933)(32.51,0.881078)(32.61,0.881223)(32.71,0.881368)(32.81,0.881512)(32.91,0.881655)(33.01,0.881798)(33.11,0.88194)(33.21,0.882082)(33.31,0.882223)(33.41,0.882364)(33.51,0.882504)(33.61,0.882643)(33.71,0.882782)(33.81,0.882921)(33.91,0.883059)(34.01,0.883196)(34.11,0.883333)(34.21,0.883469)(34.31,0.883605)(34.41,0.88374)(34.51,0.883875)(34.61,0.884009)(34.71,0.884143)(34.81,0.884276)(34.91,0.884409)(35.01,0.884541)(35.11,0.884673)(35.21,0.884805)(35.31,0.884935)(35.41,0.885066)(35.51,0.885196)(35.61,0.885325)(35.71,0.885454)(35.81,0.885582)(35.91,0.88571)(36.01,0.885838)(36.11,0.885965)(36.21,0.886092)(36.31,0.886218)(36.41,0.886343)(36.51,0.886469)(36.61,0.886593)(36.71,0.886718)(36.81,0.886842)(36.91,0.886965)(37.01,0.887088)(37.11,0.887211)(37.21,0.887333)(37.31,0.887455)(37.41,0.887576)(37.51,0.887697)(37.61,0.887817)(37.71,0.887937)(37.81,0.888057)(37.91,0.888176)(38.01,0.888295)(38.11,0.888414)(38.21,0.888532)(38.31,0.888649)(38.41,0.888766)(38.51,0.888883)(38.61,0.889)(38.71,0.889116)(38.81,0.889231)(38.91,0.889347)(39.01,0.889461)(39.11,0.889576)(39.21,0.88969)(39.31,0.889804)(39.41,0.889917)(39.51,0.89003)(39.61,0.890142)(39.71,0.890255)(39.81,0.890366)(39.91,0.890478)(40.01,0.890589)(40.11,0.8907)(40.21,0.89081)(40.31,0.89092)(40.41,0.89103)(40.51,0.891139)(40.61,0.891248)(40.71,0.891357)(40.81,0.891465)(40.91,0.891573)(41.01,0.89168)(41.11,0.891788)(41.21,0.891894)(41.31,0.892001)(41.41,0.892107)(41.51,0.892213)(41.61,0.892319)(41.71,0.892424)(41.81,0.892529)(41.91,0.892633)(42.01,0.892737)(42.11,0.892841)(42.21,0.892945)(42.31,0.893048)(42.41,0.893151)(42.51,0.893253)(42.61,0.893356)(42.71,0.893458)(42.81,0.893559)(42.91,0.893661)(43.01,0.893762)(43.11,0.893862)(43.21,0.893963)(43.31,0.894063)(43.41,0.894163)(43.51,0.894262)(43.61,0.894361)(43.71,0.89446)(43.81,0.894559)(43.91,0.894657)(44.01,0.894755)(44.11,0.894853)(44.21,0.89495)(44.31,0.895047)(44.41,0.895144)(44.51,0.895241)(44.61,0.895337)(44.71,0.895433)(44.81,0.895529)(44.91,0.895624)(45.01,0.895719)(45.11,0.895814)(45.21,0.895909)(45.31,0.896003)(45.41,0.896097)(45.51,0.896191)(45.61,0.896284)(45.71,0.896377)(45.81,0.89647)(45.91,0.896563)(46.01,0.896656)(46.11,0.896748)(46.21,0.89684)(46.31,0.896931)(46.41,0.897023)(46.51,0.897114)(46.61,0.897204)(46.71,0.897295)(46.81,0.897385)(46.91,0.897475)(47.01,0.897565)(47.11,0.897655)(47.21,0.897744)(47.31,0.897833)(47.41,0.897922)(47.51,0.898011)(47.61,0.898099)(47.71,0.898187)(47.81,0.898275)(47.91,0.898362)(48.01,0.89845)(48.11,0.898537)(48.21,0.898624)(48.31,0.89871)(48.41,0.898797)(48.51,0.898883)(48.61,0.898969)(48.71,0.899055)(48.81,0.89914)(48.91,0.899225)(49.01,0.89931)(49.11,0.899395)(49.21,0.899479)(49.31,0.899564)(49.41,0.899648)(49.51,0.899732)(49.61,0.899815)(49.71,0.899899)(49.81,0.899982)(49.91,0.900065)
};
\legend{$\varrho_\text{p}/\varrho=20$,$\varrho_\text{p}/\varrho=0.5$}
\end{axis}
\begin{axis}[anchor=south east ,xshift=6.5cm, yshift=1cm, width=0.25\textwidth, xmin=0, xmax=4, xlabel=$\sqrt{\omega/\omega_\text{f}}$, ylabel=$m\omega_0^2 C_\X(\omega)$,style={font=\scriptsize}, xlabel near ticks, ylabel near ticks]
\addplot[color=blue, dashed] coordinates {
(0., 0.)(0.05, 0.00826338)(0.1, 0.0180249)(0.15, 0.0293397)  
(0.2, 0.0422851)(0.25, 0.0569682)(0.3, 0.0735325)(0.35,0.0921645)(0.4, 0.1131)(0.45, 0.13663)(0.5, 0.163099)(0.55,0.192907)(0.6, 0.226491)(0.65, 0.264294)(0.7, 0.306704)(0.75,  
0.353936)(0.8, 0.405861)(0.85, 0.461735)(0.9, 0.519864)(0.95,  
0.577275)(1., 0.62956)(1., 0.62956)(1.01, 0.638928)(1.02, 0.647823)(1.03, 0.656198)  
(1.04, 0.664008)(1.05, 0.671207)(1.06, 0.677751)(1.07,  
0.683599)(1.08, 0.688713)(1.09, 0.69306)(1.1, 0.696608)(1.11,  
0.699331)(1.12, 0.70121)(1.13, 0.702229)(1.14, 0.702378)  
(1.15, 0.701654)(1.16, 0.70006)(1.17, 0.697603)(1.18,  
0.694299)(1.19, 0.690168)(1.2, 0.685234)(1.21, 0.679529)  
(1.22, 0.673086)(1.23, 0.665947)(1.24, 0.658152)(1.25,  
0.649746)(1.26, 0.640778)(1.27, 0.631296)(1.28, 0.62135)  
(1.29, 0.610991)(1.3, 0.600268)(1.31, 0.589231)(1.32,  
0.577928)(1.33, 0.566407)(1.34, 0.554712)(1.35, 0.542887)  
(1.36, 0.530972)(1.37, 0.519006)(1.38, 0.507024)(1.39,  
0.495059)(1.4, 0.483141)(1.41, 0.471299)(1.42, 0.459557)  
(1.43, 0.447937)(1.44, 0.43646)(1.45, 0.425143)(1.46,  
0.414002)(1.47, 0.403049)(1.48, 0.392296)(1.49, 0.381753)  
(1.5, 0.371427)(1.51, 0.361323)(1.52, 0.351448)(1.53,  
0.341804)(1.54, 0.332394)(1.55, 0.323218)(1.56, 0.314277)  
(1.57, 0.30557)(1.58, 0.297096)(1.59, 0.288853)(1.6, 0.280839)  
(1.61, 0.273049)(1.62, 0.265482)(1.63, 0.258132)(1.64,  
0.250996)(1.65, 0.24407)(1.66, 0.237349)(1.67, 0.230828)  
(1.68, 0.224502)(1.69, 0.218367)(1.7, 0.212418)(1.71,  
0.206649)(1.72, 0.201056)(1.73, 0.195634)(1.74, 0.190377)  
(1.75, 0.185281)(1.76, 0.180342)(1.77, 0.175554)(1.78,  
0.170912)(1.79, 0.166413)(1.8, 0.162052)(1.81, 0.157824)  
(1.82, 0.153725)(1.83, 0.149752)(1.84, 0.145899)(1.85,  
0.142163)(1.86, 0.138541)(1.87, 0.135028)(1.88, 0.13162)  
(1.89, 0.128316)(1.9, 0.12511)(1.91, 0.122)(1.92, 0.118983)  
(1.93, 0.116055)(1.94, 0.113213)(1.95, 0.110456)(1.96,  
0.107779)(1.97, 0.105181)(1.98, 0.102658)(1.99, 0.100208)(2.,  
0.0978292)(2., 0.0978292)(2.05, 0.0869166)(2.1, 0.0774553)  
(2.15, 0.0692254)(2.2, 0.0620434)(2.25, 0.0557556)(2.3,  
0.0502333)(2.35, 0.0453685)(2.4, 0.0410702)(2.45, 0.0372616)  
(2.5, 0.0338776)(2.55, 0.030863)(2.6, 0.0281707)(2.65,  
0.0257605)(2.7, 0.0235978)(2.75, 0.0216529)(2.8, 0.0199001)  
(2.85, 0.0183173)(2.9, 0.0168852)(2.95, 0.0155871)(3.,  
0.0144082)(3.05, 0.0133358)(3.1, 0.0123588)(3.15, 0.0114671)  
(3.2, 0.010652)(3.25, 0.00990602)(3.3, 0.0092222)(3.35,  
0.00859453)(3.4, 0.00801764)(3.45, 0.00748676)(3.5,  
0.00699761)(3.55, 0.00654639)(3.6, 0.00612967)(3.65,  
0.00574439)(3.7, 0.0053878)(3.75, 0.00505742)(3.8, 0.00475101)  
(3.85, 0.00446656)(3.9, 0.00420224)(3.95, 0.0039564)(4.,  
0.00372756)
};
\addplot[color=red, dash pattern=on 4pt off 1pt on 1pt off 1pt] coordinates {
(0., 0.)(0.01, 0.0000383723)(0.02, 0.0000782091)(0.03, 
0.000119517)(0.04, 0.000162302)(0.05, 0.000206571)(0.06, 
0.000252329)(0.07, 0.000299583)(0.08, 0.000348339)(0.09, 
0.000398603)(0.1, 0.00045038)(0.11, 0.000503677)(0.12, 
0.000558499)(0.13, 0.000614852)(0.14, 0.000672741)(0.15, 
0.000732173)(0.16, 0.000793154)(0.17, 0.00085569)(0.18, 
0.000919786)(0.19, 0.00098545)(0.2, 0.00105269)(0.21, 
0.0011215)(0.22, 0.00119191)(0.23, 0.0012639)(0.24, 0.0013375) 
(0.25, 0.00141271)(0.26, 0.00148953)(0.27, 0.00156798)(0.28, 
0.00164807)(0.29, 0.00172979)(0.3, 0.00181317)(0.31, 
0.0018982)(0.32, 0.00198491)(0.33, 0.0020733)(0.34, 
0.00216339)(0.35, 0.00225518)(0.36, 0.00234869)(0.37, 
0.00244393)(0.38, 0.00254092)(0.39, 0.00263967)(0.4, 
0.00274019)(0.41, 0.00284251)(0.42, 0.00294663)(0.43, 
0.00305259)(0.44, 0.00316039)(0.45, 0.00327005)(0.46, 
0.00338161)(0.47, 0.00349508)(0.48, 0.00361048)(0.49, 
0.00372784)(0.5, 0.00384719)(0.51, 0.00396854)(0.52, 
0.00409194)(0.53, 0.00421741)(0.54, 0.00434499)(0.55, 
0.0044747)(0.56, 0.00460658)(0.57, 0.00474067)(0.58, 
0.00487701)(0.59, 0.00501564)(0.6, 0.00515659)(0.61, 
0.00529993)(0.62, 0.00544568)(0.63, 0.00559391)(0.64, 
0.00574466)(0.65, 0.00589799)(0.66, 0.00605396)(0.67, 
0.00621261)(0.68, 0.00637403)(0.69, 0.00653827)(0.7, 
0.0067054)(0.71, 0.00687549)(0.72, 0.00704862)(0.73, 
0.00722486)(0.74, 0.00740431)(0.75, 0.00758704)(0.76, 
0.00777315)(0.77, 0.00796273)(0.78, 0.00815588)(0.79, 
0.00835271)(0.8, 0.00855332)(0.81, 0.00875784)(0.82, 
0.00896637)(0.83, 0.00917905)(0.84, 0.00939601)(0.85, 
0.00961739)(0.86, 0.00984333)(0.87, 0.010074)(0.88, 0.0103095) 
(0.89, 0.0105501)(0.9, 0.0107959)(0.91, 0.0110471)(0.92, 
0.0113038)(0.93, 0.0115664)(0.94, 0.011835)(0.95, 0.0121099) 
(0.96, 0.0123912)(0.97, 0.0126792)(0.98, 0.0129742)(0.99, 
0.0132765)(1., 0.0135864)(1.01, 0.013904)(1.02, 0.0142299) 
(1.03, 0.0145643)(1.04, 0.0149076)(1.05, 0.0152601)(1.06, 
0.0156222)(1.07, 0.0159945)(1.08, 0.0163772)(1.09, 0.016771) 
(1.1, 0.0171762)(1.11, 0.0175934)(1.12, 0.0180232)(1.13, 
0.0184661)(1.14, 0.0189229)(1.15, 0.019394)(1.16, 0.0198804) 
(1.17, 0.0203826)(1.18, 0.0209015)(1.19, 0.021438)(1.2, 
0.0219929)(1.21, 0.0225673)(1.22, 0.0231621)(1.23, 0.0237784) 
(1.24, 0.0244174)(1.25, 0.0250804)(1.26, 0.0257687)(1.27, 
0.0264837)(1.28, 0.0272271)(1.29, 0.0280003)(1.3, 0.0288053) 
(1.31, 0.0296438)(1.32, 0.0305181)(1.33, 0.0314303)(1.34, 
0.0323827)(1.35, 0.033378)(1.36, 0.034419)(1.37, 0.0355087) 
(1.38, 0.0366505)(1.39, 0.0378478)(1.4, 0.0391045)(1.41, 
0.040425)(1.42, 0.0418138)(1.43, 0.043276)(1.44, 0.044817) 
(1.45, 0.0464431)(1.46, 0.0481607)(1.47, 0.0499773)(1.48, 
0.0519009)(1.49, 0.0539405)(1.5, 0.0561058)(1.51, 0.0584079) 
(1.52, 0.0608589)(1.53, 0.0634724)(1.54, 0.0662635)(1.55, 
0.0692492)(1.56, 0.0724485)(1.57, 0.0758829)(1.58, 0.0795765) 
(1.59, 0.0835567)(1.6, 0.0878546)(1.61, 0.0925057)(1.62, 
0.0975506)(1.63, 0.103036)(1.64, 0.109015)(1.65, 0.11555) 
(1.66, 0.122714)(1.67, 0.13059)(1.68, 0.139277)(1.69, 
0.148892)(1.7, 0.159574)(1.71, 0.171486)(1.72, 0.184826) 
(1.73, 0.199835)(1.74, 0.216801)(1.75, 0.236082)(1.76, 
0.258118)(1.77, 0.283458)(1.78, 0.312792)(1.79, 0.346999) 
(1.8, 0.387206)(1.81, 0.43488)(1.82, 0.491951)(1.83, 0.56099) 
(1.84, 0.645474)(1.85, 0.750167)(1.86, 0.881689)(1.87, 
1.04935)(1.88, 1.26636)(1.89, 1.55146)(1.9, 1.93075)(1.91, 
2.43842)(1.92, 3.1117)(1.93, 3.96748)(1.94, 4.93903)(1.95, 
5.78169)(1.96, 6.10273)(1.97, 5.69733)(1.98, 4.8009)(1.99, 
3.80887)(2., 2.95295)(2.01, 2.28848)(2.02, 1.79239)(2.03, 
1.42477)(2.04, 1.15044)(2.05, 0.943019)(2.06, 0.783785)(2.07, 
0.659644)(2.08, 0.561424)(2.09, 0.482636)(2.1, 0.418633) 
(2.11, 0.366037)(2.12, 0.322358)(2.13, 0.285737)(2.14, 
0.254764)(2.15, 0.228359)(2.16, 0.205683)(2.17, 0.186081) 
(2.18, 0.16903)(2.19, 0.154115)(2.2, 0.141)(2.21, 0.129412) 
(2.22, 0.119126)(2.23, 0.109959)(2.24, 0.101757)(2.25, 
0.0943912)(2.26, 0.0877541)(2.27, 0.0817544)(2.28, 0.0763147) 
(2.29, 0.0713687)(2.3, 0.0668597)(2.31, 0.0627386)(2.32, 
0.0589631)(2.33, 0.0554965)(2.34, 0.0523067)(2.35, 0.0493657) 
(2.36, 0.0466488)(2.37, 0.0441343)(2.38, 0.0418032)(2.39, 
0.0396383)(2.4, 0.0376248)(2.41, 0.0357491)(2.42, 0.0339993) 
(2.43, 0.0323647)(2.44, 0.0308356)(2.45, 0.0294034)(2.46, 
0.0280604)(2.47, 0.0267994)(2.48, 0.0256142)(2.49, 0.0244989) 
(2.5, 0.0234484)(2.51, 0.0224579)(2.52, 0.021523)(2.53, 
0.0206398)(2.54, 0.0198047)(2.55, 0.0190144)(2.56, 0.0182658) 
(2.57, 0.0175562)(2.58, 0.016883)(2.59, 0.0162438)(2.6, 
0.0156365)(2.61, 0.015059)(2.62, 0.0145096)(2.63, 0.0139865) 
(2.64, 0.0134881)(2.65, 0.0130129)(2.66, 0.0125597)(2.67, 
0.012127)(2.68, 0.0117138)(2.69, 0.011319)(2.7, 0.0109415) 
(2.71, 0.0105803)(2.72, 0.0102347)(2.73, 0.00990365)(2.74, 
0.00958651)(2.75, 0.00928251)(2.76, 0.00899098)(2.77, 
0.00871128)(2.78, 0.0084428)(2.79, 0.00818499)(2.8, 
0.00793732)(2.81, 0.00769928)(2.82, 0.00747042)(2.83, 
0.00725029)(2.84, 0.00703848)(2.85, 0.00683459)(2.86, 
0.00663827)(2.87, 0.00644915)(2.88, 0.00626692)(2.89, 
0.00609125)(2.9, 0.00592186)(2.91, 0.00575848)(2.92, 
0.00560082)(2.93, 0.00544866)(2.94, 0.00530174)(2.95, 
0.00515986)(2.96, 0.00502278)(2.97, 0.00489032)(2.98, 
0.00476228)(2.99, 0.00463848)(3., 0.00451874)(3.01, 
0.00440291)(3.02, 0.00429082)(3.03, 0.00418233)(3.04, 
0.00407729)(3.05, 0.00397557)(3.06, 0.00387704)(3.07, 
0.00378158)(3.08, 0.00368906)(3.09, 0.00359939)(3.1, 
0.00351245)(3.11, 0.00342813)(3.12, 0.00334635)(3.13, 
0.00326701)(3.14, 0.00319002)(3.15, 0.0031153)(3.16, 
0.00304275)(3.17, 0.00297232)(3.18, 0.00290391)(3.19, 
0.00283747)(3.2, 0.00277292)(3.21, 0.00271019)(3.22, 
0.00264922)(3.23, 0.00258996)(3.24, 0.00253235)(3.25, 
0.00247632)(3.26, 0.00242183)(3.27, 0.00236882)(3.28, 
0.00231725)(3.29, 0.00226707)(3.3, 0.00221823)(3.31, 
0.00217069)(3.32, 0.00212441)(3.33, 0.00207934)(3.34, 
0.00203546)(3.35, 0.00199271)(3.36, 0.00195107)(3.37, 
0.00191051)(3.38, 0.00187098)(3.39, 0.00183245)(3.4, 
0.0017949)(3.41, 0.0017583)(3.42, 0.00172261)(3.43, 
0.00168781)(3.44, 0.00165388)(3.45, 0.00162078)(3.46, 
0.00158849)(3.47, 0.00155699)(3.48, 0.00152625)(3.49, 
0.00149626)(3.5, 0.00146698)(3.51, 0.00143841)(3.52, 
0.00141051)(3.53, 0.00138328)(3.54, 0.00135669)(3.55, 
0.00133072)(3.56, 0.00130536)(3.57, 0.00128058)(3.58, 
0.00125638)(3.59, 0.00123274)(3.6, 0.00120964)(3.61, 
0.00118706)(3.62, 0.001165)(3.63, 0.00114343)(3.64, 
0.00112235)(3.65, 0.00110174)(3.66, 0.0010816)(3.67, 
0.00106189)(3.68, 0.00104263)(3.69, 0.00102379)(3.7, 
0.00100536)(3.71, 0.000987329)(3.72, 0.000969693)(3.73, 
0.000952439)(3.74, 0.000935556)(3.75, 0.000919034)(3.76, 
0.000902866)(3.77, 0.000887042)(3.78, 0.000871553)(3.79, 
0.00085639)(3.8, 0.000841546)(3.81, 0.000827013)(3.82, 
0.000812783)(3.83, 0.000798848)(3.84, 0.000785201)(3.85, 
0.000771835)(3.86, 0.000758744)(3.87, 0.00074592)(3.88, 
0.000733357)(3.89, 0.000721049)(3.9, 0.000708989)(3.91, 
0.000697172)(3.92, 0.000685592)(3.93, 0.000674244)(3.94, 
0.000663121)(3.95, 0.000652219)(3.96, 0.000641532)(3.97, 
0.000631056)(3.98, 0.000620785)(3.99, 0.000610714)(4., 
0.00060084)
};
\end{axis}
\end{tikzpicture}
\caption{The effective temperature $T^\X$ governing the Boltzmann
  factor of a hot Brownian particle in harmonic confinement (normalized to the 
  temperature difference $\Delta T$ between the particle surface
  and the ambient temperature $T_0$). The
  dashed-dotted and dashed lines correspond to weak
  ($\varrho_\text{p}/\varrho=20$), and strong
  ($\varrho_\text{p}/\varrho=0.5$) coupling to the solvent, or
  under- and over-damped oscillations, respectively. 
  For large but physically accessible values of the
  particle/fluid density ratio $\varrho_\text{p}/\varrho$, 
  $T^\X$ coincides with the noise temperature
  $\caT(\omega_0)$ (solid line), in agreement with
  Eq.~(\ref{Energy}). Under these conditions, the Brownian particle
  can serve as a thermospectrometer for the noise 
  spectrum $\kB \caT(\omega)$. \emph{Inset:} the corresponding position spectral densities evaluated at $\omega_0/\omega_\text{f}=4$.}
\label{fig:Teff}
\end{figure}

\emph{Brownian thermospectrometry}---The (at first sight) maybe somewhat disturbing dependence of the
apparent equipartition temperatures on the confinement can actually be exploited to
restore Onsager regression and the classical FDT, albeit with an
effective temperature.  If the motion of the Brownian particle is only
weakly damped by the solvent, the response function is sharply peaked
around the eigenfrequency $\omega_0$. This limit can practically be
realized for large particle-to-fluid density ratios $\varrho_\text{p}/\varrho\gg1$,
e.g.\ for a Brownian particle suspended in a gas
\cite{Raizen.2010,Li.2011,Millen.2014}\footnote{Ultimately, for a rarified gas
  in the Knudsen regime, our local equilibrium assumption breaks down. The effective Brownian temperature
  can then be estimated using kinetic theory \cite{Millen.2014}.}. Then almost
only the resonant mode $\kB \caT(\omega_0)$ of the thermal energy
spectrum is on average absorbed (Fig.~\ref{fig:Teff}). Thereby, from the Fourier back-transform of
Eq.~\eqref{Cx}, the mean energy follows in the familiar equipartition form:
\begin{equation}\label{Energy}
 m\omega_0^2 \mean{X_i^2}= M\mean{V_i^2}=\kB \caT(\omega_0)\,.
\end{equation}
This shows that by varying the trap stiffness $m\omega_0^2$, i.e.\ by
endowing the Brownian thermometer with a tunable frequency filter, one
can turn it into a genuine thermospectrometer. It then measures the
energy content of the position and velocity coordinates $X_i$ and
$V_i$ at a prescribed frequency $\omega_0$. Since each Fourier mode is thermalized at its
own temperature $\caT(\omega_0)$ one naturally concludes that the
noise temperature acts as an effective temperature for such ``tuned''
Brownian particles. Moreover, Eq.~\eqref{Cx} can be approximated at each
(positive) time by
\begin{equation}
  \dot C_\X(t) = - \kB\caT(\omega_0)R_\X(t) \, ,
\end{equation}
which evidently is the classical FDT. So the weak-coupling limit
restores the classical FDT in the non-equilibrium system as much as 
it does in the quantum case \cite{Ford.1999}.

\emph{Discussion}---It seems interesting to note that a quantity completely
analogous to our $\caT(\omega)$, endowed with the very same physical
meaning, is commonly used in electronics. In that context, the
noise temperature is introduced to account for fluctuations in
non-equilibrium conductors, when the Johnson--Nyquist 
FDT in not satisfied  \cite{skolnik}. Exactly as in the equilibrium case, the
mapping between the two related phenomena is established by
substituting in Eq.~\eqref{noise}---the analogue of Nyquist's formula
\cite{callen.1951}---the thermal force with the voltage and the friction
coefficient with the resistance. Exploiting the analogy further, one
finds that the Brownian noise temperature \eqref{Tdef} exhibits the
same formal structure as the noise temperature of radio
receivers. Just like Eq.~\eqref{Tdef}, the effective antenna
temperature, which results from the electromagnetically mediated
coupling to a non-isothermal environment, is an average over the
temperatures of the surrounding radiation field 
weighted by the radiation pattern \cite{milligan}.

The notion of a frequency-dependent temperature that quantifies violations of the FDT  far from equilibrium is moreover reminiscent of the effective temperature previously suggested to govern the linear response of glasses \cite{Cugliandolo.1997}. This idea has been tested, with mixed success, in several models \cite{Cugliandolo.2011}.
We therefore emphasize that our Eq.~(\ref{Tdef}) and the corresponding generalized
fluctuation-dissipation relations are not postulated, but analytically
derived \cite{Falasco.2014, SI}, thereby providing an independent test-bed for
rigorously analyzing the scope of the notion of effective
temperatures, far from equilibrium. Moreover, our theory may serve as
a starting point to consistently extend the notions of stochastic
thermodynamics  \cite{Seifert.2012} to non-isothermal systems.

\bibliography{short_paper}
\bibliographystyle{apsrev4-1}

\end{document}


\title{Supplemental Material}

\author{G.~Falasco, M.~V.~Gnann, K.~Kroy}

\maketitle

We consider a particle of arbitrary shape immersed in a non-isothermal simple fluid in local thermal equilibrium. The dynamical state of the system is given is terms of a few reduced variables, evolving in accord with the linearized fluctuating hydrodynamic equations and Newton's law. The particle is described by the center of mass $\vec X(t)$, and the translational and rotational velocity, respectively, $\vec V(t)$ and $\vec \Omega (t)$. The fluid occupies the volume $\caV$ and is described by the velocity and temperature field, respectively, $\vec v(\vec r,t)$ and $T(\vec r,t)$.
The velocity field obeys the equations
\begin{subequations}\label{sys}
\begin{align}
&\varrho \partial_t \vec v(\vec r, t) - \nabla \cdot \tens \sigma (\vec r, t) = \nabla \cdot \tens \tau (\vec r, t)\,,\label{sys1}\\
&\nabla \cdot \vec v (\vec r, t) = 0. \label{sys2}\\
& \vec v(\vec r, t) = \vec V(t) + \vec \Omega(t) \times \vec r \quad \mbox{on } \caS \label{sys3}
\end{align}
\end{subequations}
The symmetric stress tensor $\vec \sigma$ has components $\sigma_{ij}= -p \delta_{ij} +2 \eta \Gamma_{ij}$, where $p$ is the pressure and $\Gamma_{ij}= (\partial_i v_j + \partial_j v_i)/2$ the shear rate tensor. The dynamic viscosity $\eta$ is in general a function of the temperature. Hence it is space dependent, $\eta= \eta[T(\vec r)]$. For simplicity, the fluid is assumed incompressible. Therefore the mass conservation reduces to Eq.~\eqref{sys2}. We denote by $\varrho$ the constant density of the fluid. 
The boundary condition \eqref{sys3} simply states that the fluid adheres at the surface $\caS$ of the Brownian particle (no slip).

The noise stress tensor $\tens{\tau}$ describes thermal fluctuations in the fluid, whose statistical properties follow from the hypothesis of local equilibrium. Namely, $\vec \tau$ is Gaussian distributed with zero mean and second moments obeying the local fluctuation-dissipation theorem
\begin{align}\label{fluidfdt}
\left< \tau_{ij}(\vec r, t) \tau_{kl}(\vec r^\prime, t^\prime) \right>\! &=\!2 \eta(\vec r, t) \kB T(\vec r, t) \delta(\vec r - \vec r^\prime) \delta (t - t^\prime) \left(\delta_{ik} \delta_{jl} + \delta_{il} \delta_{jk} \right).\end{align}
We assume that $\tens \tau \equiv 0$ on the boundary $\caS$. The temperature field entering Eq.~\eqref{fluidfdt} is in general the solution of a heat equation with appropriate boundary conditions describing the heat sources in $\caV$ and the fluxes across $\caS$. In the following we assume the temperature field to be a prescribed function, independent of time. This is indeed the case whenever the particle motion causes negligible disturbances to the fluid temperature. 

The velocities evolve by Newton's equations of motion, i.e.
\begin{subequations}\label{newton}
\begin{eqnarray}
m \dot{\vec V}(t) &=& \vec F(t) + \vec{F}_{\text{e}}(t) , \label{newton1}\\
\tens{I} \cdot \dot{\vec \Omega}(t) &=& \vec T(t) +\vec{T}_{\text{e}}(t) \label{newton2}
\end{eqnarray}
\end{subequations}
where $m$ is the mass of the particle and $\tens I$ its tensor of inertia. The force and torque exerted by the fluid are
\begin{subequations}\label{formom}
\begin{eqnarray}
\vec F(t) &=&- \int_{\caS} \tens \sigma (\vec r, t) \cdot \vec n(\vec r) \, \diff^2 r,\\
\vec T(t) &=& - \int_{\caS} \vec r \times \left(\tens \sigma (\vec r, t) \cdot \vec n(\vec r)\right) \diff^2 r,
\end{eqnarray}
\end{subequations}
with $\vec n(\vec r)$ the inner normal vector field of the particle surface $\caS$. External forces  $\vec{F}_{\text{e}}(t)$ and torques $\vec{T}_{\text{e}}(t)$ may also be present. 
The system \eqref{sys}--\eqref{formom} describes the evolution of the fluid and the Brownian particle entirely.

Our aim is to eliminate the hydrodynamic fields and reduce Eqs. \eqref{sys}--\eqref{formom} to a generalized Langevin equation for the particle variables only. Thus we rewrite system \eqref{newton} in the form
\begin{equation}\label{newton3}
\tens L \cdot \dot{\vec b}(t) = \vec h (t) + \vec{f}_\text{e}(t),
\end{equation}
where we combine the translational and rotational velocity into the $6$-vector $\vec b(t) \equiv \left(\vec V(t), \vec \Omega(t)\right)$, and we define the generalized tensor of inertia
\[
\tens L \equiv \begin{pmatrix} m  & 0\\ 0 & \tens I \end{pmatrix},
\]
and the generalized forces
\begin{equation}\label{newton4}
\vec h(t) \equiv \begin{pmatrix} \vec F(t) \\ \vec T(t) \end{pmatrix}, \quad \vec{f}_\text{e}(t) \equiv \begin{pmatrix} \vec{F}_\text{e}(t) \\ \vec{T}_\text{e}(t) \end{pmatrix}.
\end{equation}
One should appreciate that the hydrodynamic force splits into two parts
\begin{equation}\label{split}
\vec h(t) \equiv \vec{h}_\text{r}(t) + \vec{\xi}(t),
\end{equation}
where $\vec{h}_\text{r}(t)$ is identified with the \emph{resistance} force and $\vec{\xi}(t)$ with the \emph{Langevin noise}.
Equation \eqref{split} is easily justified as follows. By equation \eqref{sys}, $\vec u(\vec r, t)$ and $ p(\vec r, t)$ are linear functionals of $\vec b(t^\prime)$ with $- \infty < t^\prime < t$. Therefore, in view of Eq.~\eqref{formom}, the total hydrodynamic force necessarily contains a contribution which is a linear functional of $\vec b(t^\prime)$ with $- \infty < t^\prime < t$, i.e. we can write
\begin{equation}\label{sysforce}
\vec h_\text{r}(t) = - \int_{-\infty}^t \tens Z(\vec X, t - t^\prime) \cdot \vec b(t^\prime) \diff t^\prime.
\end{equation}
Here $\tens Z(\vec X, t)$ is a $6 \times 6$ time-dependent friction tensor, which also depends on the particle position owing to the non-constant fluid viscosity. We omit to show this dependence in the following. Moreover, since \eqref{sys1} is a linear but non-homogeneous equation, the term $\vec \xi(t)$ has to be included in Eq.~\eqref{split} in order to account for contributions to the hydrodynamic force which are independent of the particle velocity.

In the subsequent derivation we shall derive the statistics of the Langevin noise $\vec \xi(t)$ and relate it to the dissipative term $\vec h_\text{r}(t)$. The linearity of the problem suggests to operate in the frequency space. Given a generic function of time $g(t)$, we denote its Fourier transform and half-Fourier transform, respectively, by
\begin{align*}
&g(\omega)=\int_{-\infty}^{\infty} g(t) e^{-i \omega t} \diff t, &&g^{\+}(\omega)=\int_{0}^{\infty} g(t) e^{-i \omega t} \diff t.
\end{align*}
The complex conjugate of $g(\omega)$ will be denoted by $g^*(\omega)$.

The Fourier transform of Newton's equation \eqref{newton3}, with the definitions \eqref{split} and \eqref{sysforce}, reads
\begin{equation}\label{newton3_bis}
-i \omega \tens L \cdot \vec b(\omega) = - \tens Z^{+}(\omega) \cdot  \vec b(\omega) + \vec \xi(\omega)+ {\vec f}_{\text{e}}(\omega).
\end{equation}
If the velocity 6-vector is split into its average and random part, i.e. $\vec b(\omega)=\vec {\tilde b}(\omega)+ \langle\vec b(\omega)\rangle$ the total hydrodynamic force takes the form $\vec h(\omega)= \mean{\vec{h}_\text{r}(\omega) }+\vec{\tilde h}(\omega)$, with
\begin{subequations}\label{hforce}
\begin{eqnarray}
\mean{\vec{h}_\text{r}(\omega) }&=&- \tens Z^{+}(\omega) \cdot \langle \vec b(\omega)\rangle, \label{hforce1}\\
\vec {\tilde h}(\omega) &=&- \tens Z^{+}(\omega) \cdot \vec {\tilde b}(\omega) + \vec \xi(\omega).\label{hforce2}\,
\end{eqnarray}
\end{subequations}
The physical meaning of Eqs.~\eqref{hforce} is best understood if we cast the Fourier transform of Eqs.~\eqref{sys} into the decoupled set of equations
\begin{subequations}\label{sys_bis}
\begin{align}
&-i \omega \varrho  \vec {u}(\vec r, \omega) - \nabla \cdot \tens {\sigma} (\vec r, \omega) = 0,\label{sys_bis1}\\
&\nabla \cdot \vec {u} (\vec r, \omega) = 0 \quad,\label{sys_bis2}\\
&\vec {u}(\vec r, \omega) = \langle\vec U(\omega)\rangle +\langle \vec \Omega(\omega) \rangle\times \vec r \quad \mbox{on } \caS, \label{sys_bis3}
\end{align}
\end{subequations}
and
\begin{subequations}\label{rand_bis}
\begin{align}
&-i \omega \varrho \vec {\tilde u}(\vec r, \omega) - \nabla\! \cdot \tens {\tilde \sigma} (\vec r, \omega) = \nabla \!\cdot \tens \tau (\vec r, \omega),\label{rand_bis1}\\
&\nabla \cdot \vec {\tilde u} (\vec r, \omega) = 0 \quad,\label{rand_bis2}\\
&\vec {\tilde u}(\vec r, \omega) = \vec {\tilde V}(\omega) + \vec {\tilde \Omega}(\omega) \times \vec r \quad \mbox{on } \caS. \label{rand_bis3}
\end{align}
\end{subequations}
Clearly, equations \eqref{sys_bis} describe the \emph{deterministic} fluid velocity, i.e. $\vec{ u}(\vec r, \omega)\equiv \langle\vec v(\vec r, \omega)\rangle$, while equations \eqref{rand_bis} account for its \emph{random} thermal fluctuations, i.e $ \vec{ \tilde{u}}(\vec r, \omega)\equiv\vec v(\vec r,t)- \vec{ u}(\vec r, \omega)$. In view of the boundary conditions \eqref{sys_bis3} and \eqref{rand_bis3}, one recognizes \eqref{hforce1}  as the friction produced by the deterministic velocity field solution of Eqs. \eqref{sys_bis}, and \eqref{hforce2} as the random force exerted by the velocity field solution of Eqs. \eqref{rand_bis}. Note that the deterministic part of the velocity vector $\mean{\vec b(\omega)}$ can be chosen arbitrarily.

We are now in the position to evaluate the statistics of the Langevin noise $\vec \xi(\omega)$. We operate in three steps. First, we derive an expression for the friction tensor $\vec Z(\omega)$. Then we show that $\vec \xi(\omega)$ is a Gaussian variable with zero mean. Finally we link the noise correlation tensor $\langle \xi_i(\omega) \xi_j^*(\omega) \rangle$ to the friction tensor. The benefit coming from casting Eqs.~\eqref{sys} into Eqs.~\eqref{sys_bis}--\eqref{rand_bis} is easily seen. Indeed  one can conveniently choose the boundary conditions in Eq.~\eqref{sys_bis} to express the friction tensor and noise statistics in terms of functions of $\vec u(\vec r, \omega)$, without solving the much more involved problem represented by the stochastic equations \eqref{rand_bis}.

To evaluate the components of the friction tensor $Z_{ij}$ we make use of its properties, hinging only on the symmetry of the stress tensor $\tens \sigma$  \cite{Hauge.1973}. It can be showed that in general $\tens Z$ is symmetric and is a real even function of time. From the latter property it immediately follows the relation
\begin{equation}\label{symmetry}
Z_{ij}(\omega)=Z_{ij}^\+(\omega)+{Z_{ij}^{\+}}^*(\omega).
\end{equation}
We exploit the freedom of choosing the boundary condition \eqref{rand_bis3} to select velocity vectors whose $\alpha$-th entry is the only non-zero one, and denote them by $\tensor[^{\scriptscriptstyle \alpha}]{b}{_i}(\omega)$---the superscript $\alpha$ will also be appended to the corresponding hydrodynamic fields. 

We wish to find an expression in terms of the solution to Eqs.~\eqref{sys_bis} for the quantity
\begin{equation}\label{SymmetryZ}
Z_{ij}(\omega)\langle ^{\scriptscriptstyle \alpha}b_i(\omega)\rangle \langle ^{\scriptscriptstyle \beta}b_j^*(\omega)\rangle= Z_{\alpha \beta}(\omega)\langle b_\alpha(\omega)\rangle \langle b_\beta^*(\omega)\rangle,
\end{equation}
where the equality holds by virtue of the choice of $\vec b(\omega)$. In Eq.~\eqref{SymmetryZ} and in the following we apply the Einstein summation convention to latin indices only. Also, we suppress the function arguments where there is no risk of confusion. Equation \eqref{SymmetryZ} reads
\begin{eqnarray}\nonumber
&& Z_{ij} \mean{ ^{\scriptscriptstyle \alpha}b_i} \mean{^{\scriptscriptstyle \beta}b^*_j}\stackrel{\eqref{symmetry}}{=} (Z_{ij}^\+ +{Z_{ij}^{\+}}^*)\mean{ ^{\scriptscriptstyle \alpha}b_i} \mean{^{\scriptscriptstyle \beta}b_j^*} \stackrel{\eqref{hforce1}}{=}-\left(^{\scriptscriptstyle \alpha}{h_\text{r}}_i\mean{^{\scriptscriptstyle \beta}b_i^*}+ \mean{ ^{\scriptscriptstyle \alpha}b_i} \phantom{\,}^{\scriptscriptstyle \beta}{h_\text{r}}_i^*\right)\label{Zsym} \\
&\stackrel{\eqref{formom},\eqref{newton4}}{=}& \langle ^{\scriptscriptstyle \beta}V_i^* \rangle  \int_{\caS} \,^{\scriptscriptstyle \alpha}\sigma_{ij} n_j \, \diff^2 r + \langle ^{\scriptscriptstyle \beta}\Omega_i^* \rangle \int_{\caS} (\vec r \times \left(^{\scriptscriptstyle \alpha}\tens \sigma \cdot \vec n\right))_i \, \diff^2 r + \langle ^{\scriptscriptstyle \alpha}V_i \rangle  \int_{\caS} \,^{\scriptscriptstyle \beta}\sigma^*_{ij} n_j\, \diff^2 r + \langle ^{\scriptscriptstyle \alpha}\Omega_i \rangle \int_{\caS} (\vec r \times \left(^{\scriptscriptstyle \beta}\tens{\sigma}^* \cdot \vec n\right))_i\,  \diff^2 r \nonumber \\
&=& \int_{\caS} \left(\langle ^{\scriptscriptstyle \beta}\vec V^* \rangle+ \langle ^{\scriptscriptstyle \beta}\vec \Omega^*\rangle \times \vec r\right)_i \! \,^{\scriptscriptstyle \alpha}\!\sigma_{ij} n_j \, \diff^2 r +\int_{\caS} \left(\langle ^{\scriptscriptstyle \alpha}\vec V \rangle+ \langle ^{\scriptscriptstyle \alpha}\vec \Omega\rangle \times \vec r\right)_i \! \,^{\scriptscriptstyle \beta}\!\sigma_{ij}^* n_j \, \diff^2 r \nonumber \\
&\stackrel{\eqref{sys_bis3}}{=}&\int_{\caS} {}^{\scriptscriptstyle \beta}u_i^* {}^{\scriptscriptstyle \alpha}\!\sigma_{ij} n_j \diff^2 r+\int_{\caS} {}^{\scriptscriptstyle \alpha}u_i {}^{\scriptscriptstyle \beta}\!\sigma_{ij}^* n_j\, \diff^2 r \nonumber \\
&=&\int_{\caV} \partial_j\left(^{\scriptscriptstyle \beta}u_i^* {}^{\scriptscriptstyle \alpha}\!\sigma_{ij}\right) \, \diff^3 r \label{div_th} +\int_{\caV} \partial_j\left(^{\scriptscriptstyle \alpha}u_i  {}^{\scriptscriptstyle \beta}\!\sigma_{ij}^*\right) \, \diff^3 r \label{divergence}   \\
&\stackrel{\eqref{sys_bis1}}{=}& \int_{\caV} \left( ^{\scriptscriptstyle \alpha}\!\sigma_{ij} \partial_j {}^{\scriptscriptstyle \beta}u_i^*+ {}^{\scriptscriptstyle \beta}\!\sigma_{ij}^* \partial_j {}^{\scriptscriptstyle \alpha}u_i\right) + i \varrho\omega \int_{\caV}  \left( ^{\scriptscriptstyle \alpha}u^*_i {}^{\scriptscriptstyle \beta}u_i- {}^{\scriptscriptstyle \beta}u^*_i {}^{\scriptscriptstyle \alpha}u_i\right) \, \diff^3 r \nonumber\\
&\stackrel{\eqref{sys_bis2}}{=}& \int_{\caV} 2 \eta \left( {}^{\scriptscriptstyle \alpha}\Gamma_{ij} \partial_j {}^{\scriptscriptstyle \beta}u_i^*+ {}^{\scriptscriptstyle \beta}\Gamma_{ij}^* \partial_j {}^{\scriptscriptstyle \alpha}u_i\right) \, \diff^3 r + i \varrho\omega \int_{\caV}  \left( ^{\scriptscriptstyle \alpha}u^*_i {}^{\scriptscriptstyle \beta}u_i- {}^{\scriptscriptstyle \beta}u^*_i {}^{\scriptscriptstyle \alpha}u_i\right) \, \diff^3 r\nonumber\\
&=& 2 \int_{\caV} \Phi_{\alpha \beta}\,\diff^3 r + i \varrho\omega \int_{\caV}  \left( ^{\scriptscriptstyle \alpha}u^*_i {}^{\scriptscriptstyle \beta}u_i- {}^{\scriptscriptstyle \beta}u^*_i {}^{\scriptscriptstyle \alpha}u_i\right) \, \diff^3 r\label{friction},
\end{eqnarray}
where in \eqref{divergence} we used the divergence theorem and in \eqref{friction} we defined the response tensor as
\begin{equation}\label{phi}
\Phi_{\alpha \beta}(\vec X, \vec r, \omega)\equiv  \eta(\vec r) \left(\partial_i {}^{\scriptscriptstyle \alpha}u_j(\vec r, \omega) \partial_i {}^{\scriptscriptstyle \beta}u_j^*(\vec r, \omega) + \partial_i {}^{\scriptscriptstyle \beta}u_j(\vec r,\omega)\partial_j {}^{\scriptscriptstyle \alpha}u_i^*(\vec r, \omega)\right)=\Phi^*_{\beta\alpha}(\vec X, \vec r, \omega),
\end{equation}
where the $\vec X$ dependence of the hydrodynamic fields is not explicitly showed.
Equation \eqref{friction} is valid whatever the magnitude of $b_\alpha$ and $b_\beta$, in particular when they are unit vectors. With this choice we have
\begin{equation}\label{cancel}
Z_{\alpha \beta}(\vec X, \omega)= 2 \int_{\caV} \Phi_{\alpha \beta} \,\diff^3 r + i \varrho\omega \int_{\caV}  \left( ^{\scriptscriptstyle \alpha}u^*_i {}^{\scriptscriptstyle \beta}u_i- {}^{\scriptscriptstyle \beta}u^*_i {}^{\scriptscriptstyle \alpha}u_i\right) \, \diff^3 r
\end{equation}
which has to be invariant under exchange of $\alpha$ and $\beta$, owing to the symmetry of $\vec Z$. Therefore one can eliminate the second term in Eq.~\eqref{cancel} and obtain for the friction tensor
\begin{equation}\label{Zetabb}
Z_{\alpha \beta}(\vec X, \omega)= \int_{\caV} \Phi_{\alpha \beta}(\vec X, \vec r, \omega) \,\diff^3 r + \int_{\caV} \Phi_{\beta \alpha}(\vec X, \vec r, \omega) \,\diff^3 r =2 \Real   \int_{\caV}\Phi_{\alpha \beta}(\vec X, \vec r, \omega) \, \diff^3 r,
\end{equation}
that can be written as
\begin{equation}\label{Zfinal}
Z_{\alpha \beta}(\vec X, \omega)= \int_{\caV}\phi_{\alpha \beta}(\vec X, \vec r, \omega) \, \diff^3 r,
\end{equation}
where $\phi_{\alpha \beta}(\vec X, \vec r, \omega)\equiv 2 \Real  \Phi_{\alpha \beta}(\vec X, \vec r ,\omega)$ is twice the real part of the response tensor \eqref{phi}, computed by setting unit vectors in Eq.~\eqref{sys_bis3}.

Then we turn to the random force $\vec \xi(\omega)$ 
\begin{eqnarray}\nonumber
&&\xi_i \langle {}^{\scriptscriptstyle \alpha}b_i \rangle \stackrel{\eqref{hforce2}}{=}\left(\tilde h_i+Z_{ij}^+ \tilde b_j \right) \langle {}^{\scriptscriptstyle \alpha}b_i\rangle \stackrel{\eqref{hforce1}}{=} \tilde h_i\langle {}^{\scriptscriptstyle \alpha}b_i \rangle- {}^{\scriptscriptstyle \alpha}h_j \tilde b_j \nonumber \\
&\stackrel{\eqref{formom}}{=}& -\langle {}^{\scriptscriptstyle \alpha}V_i\rangle  \int_{\caS} \tilde \sigma_{ij} n_j \, \diff^2 r - \langle {}^{\scriptscriptstyle \alpha}\Omega_i\rangle \int_{\caS} (\vec r \times \left( \tens{\tilde \sigma} \cdot \vec n\right))_i  \,\diff^2 r +\tilde V_i  \int_{\caS} {}^{\scriptscriptstyle \alpha}\!\sigma_{ij} n_j \, \diff^2 r +\tilde\Omega_i \int_{\caS} (\vec r \times \left(\tens {}^{\scriptscriptstyle \alpha}\!\sigma \cdot \vec n\right))_i  \,\diff^2 r \nonumber\\
&=& -\int_{\caS} \big(\langle^{\scriptscriptstyle \alpha}\vec V \rangle + \langle {}^{\scriptscriptstyle \alpha} \vec \Omega \rangle \times \vec r\big)_i \tilde \sigma_{ij}n_j \, \diff^2 r +\int_{\caS} \big(\vec {\tilde V}+ \vec {\tilde \Omega} \times \vec r\big)_i {}^{\scriptscriptstyle \alpha}\sigma_{ij} n_j \, \diff^2 r \nonumber \\
&\stackrel{\eqref{sys_bis3},\eqref{rand_bis3} }{=}&-\int_{\caS} {}^{\scriptscriptstyle \alpha}u_i \tilde \sigma_{ij} n_j\, \diff^2 r +\int_{\caS} \tilde u_i {}^{\scriptscriptstyle \alpha}\sigma_{ij} n_j \, \diff^2 r \nonumber \\
&=& -\int_{\caV} \partial_j\left(  \tilde u_i {}^{\scriptscriptstyle \alpha}\!\sigma_{ij}\right)\, \diff^3 r + \int_{\caV} \partial_j\left(  {}^{\scriptscriptstyle \alpha}u_i \tilde \sigma_{ij}\right)\, \diff^3 r  \label{div_th2} \\
&\stackrel{\eqref{sys_bis1}, \eqref{rand_bis1}}{=}&-\int_{\caV} {}^{\scriptscriptstyle \alpha}\sigma_{ij} \partial_j \tilde u_i \diff^3 r- \int_{\caV} {}^{\scriptscriptstyle \alpha}u_i \partial_j  \tau_{ij} \diff^3 r +\int_{\caV} \tilde \sigma_{ij}\partial_j  {}^{\scriptscriptstyle \alpha}u_i \diff^3 r \label{trick} \\
&\stackrel{\eqref{sys_bis2}, \eqref{rand_bis2}}{=}& -\int_{\caV} {}^{\scriptscriptstyle \alpha}u_i \partial_j \tau_{ij} \diff^3 r \nonumber \\
&=& \int_{\caV}  \tau_{ij}  \partial_j {}^{\scriptscriptstyle \alpha}u_i \diff^3 r \nonumber.
\end{eqnarray}
In \eqref{div_th2} we made use of the divergence theorem, and in \eqref{trick} of the property $ \tilde\sigma_{ij} \partial_j  {}^{\scriptscriptstyle \alpha}u_i ={}^{\scriptscriptstyle \alpha}\sigma_{ij} \partial_j \tilde u_i $, which is a direct consequence of the symmetry of $\vec \sigma$.
We have found
\begin{equation}\label{noise}
\xi_i \langle {}^{\scriptscriptstyle \alpha}b_i \rangle = \xi_\alpha \langle b_\alpha\rangle=\int_{\caV} \tau_{ij}  \partial_j {}^{\scriptscriptstyle \alpha}u_i\,\diff^3 r,
\end{equation}
which shows that $\vec \xi$ is Gaussian with vanishing mean, being the integral of the deterministic quantity $\partial_j {}^{\scriptscriptstyle \alpha}u_i$ times the zero-mean Gaussian field $\tens \tau$. Hence, its correlation matrix suffices to specify the statistics  completely. In addition, $\vec \xi$ inherits the time reversal invariance of the stress tensor $\tens \tau$.

Using \eqref{noise}, we determine the noise correlations 
\begin{eqnarray}
&&\langle \xi_i(\omega)  \xi_j^*(\omega')\rangle \langle {}^{\scriptscriptstyle \alpha}b_i(\omega)\rangle \langle {}^{\scriptscriptstyle \beta}b_j^*(\omega')\rangle = \langle \xi_\alpha(\omega)  \xi_\beta^*(\omega')\rangle \langle b_\alpha(\omega)\rangle \langle b_\beta^*(\omega')\rangle \label{xixibb}\\
&=& \int_{\caV} \diff^3 r' \int_{\caV} \diff^3 r  \, \partial_j {}^{\scriptscriptstyle \alpha}u_i(\vec r,\omega)  \langle  \tau_{ij}(\vec r, \omega) \tau_{kl}^*(\vec r', \omega') \rangle \partial_l {}^{\scriptscriptstyle \beta}u_k^*(\vec r',\omega') \nonumber\\
&=& 2 \kB \delta(\omega-\omega') \int_{\caV} \eta(\vec r) T(\vec r) \left(\partial_i {}^{\scriptscriptstyle \alpha}u_j(\vec r, \omega) \partial_i {}^{\scriptscriptstyle \beta}u_j^*(\vec r, \omega) + \partial_i {}^{\scriptscriptstyle \alpha}u_j(\vec r,\omega)\partial_j {}^{\scriptscriptstyle \beta}u_i^*(\vec r, \omega)\right) \, \diff^3 r \label{noisefdt} \\
&=& 2 \kB \delta(\omega-\omega') \int_{\caV}  \Phi_{\alpha \beta}(\vec X, \vec r, \omega) T(\vec r) \, \diff^3 r \label{corr}.
\end{eqnarray}
In \eqref{noisefdt} we used the Fourier transform of \eqref{fluidfdt}. One gets, setting the magnitude of $\langle b_\alpha\rangle$ and $\langle b_\beta^*\rangle$ to one
\begin{equation}\label{xixi}
\langle \xi_\alpha(\vec X, \omega)   \xi_\beta^*(\vec X, \omega')\rangle= 2 \kB \delta(\omega-\omega') \int_{\caV}  \Phi_{\alpha \beta}(\vec X, \vec r, \omega) T(\vec r) \, \diff^3 r.
\end{equation}
Besides, the symmetry under time reversal of $\vec \xi$ implies that Eq.~\eqref{xixi} is invariant under complex conjugation. This yields immediately
\begin{align}
&\langle \xi_\alpha(\vec X, \omega)  \xi_\beta^*(\vec X, \omega')\rangle=\kB \delta(\omega-\omega') 2 \Real  \int_{\caV}  \Phi_{\alpha \beta}(\vec X, \vec r, \omega) T(\vec r) \, \diff^3 r. \label{xifinal}
\end{align} 

From Eqs.~\eqref{Zfinal} and \eqref{xifinal} we can finally obtain the noise correlation tensor in the form
\begin{equation}\label{corr}
\langle \xi_\alpha(\vec X, \omega)   \xi_\beta^*(\vec X, \omega')\rangle= \kB \caT_{\alpha \beta}(\vec X, \omega) Z_{\alpha \beta}(\vec X, \omega) \delta(\omega-\omega'),
\end{equation}
where $\caT_{\alpha\beta}$ is the frequency-dependent noise temperature defined by the spatial average of the temperature field $T(\vec r)$ performed with the response tensor $\phi_{\alpha \beta}$
\begin{equation}\label{Tdef}
\caT_{\alpha\beta}(\vec X,\omega)\equiv\frac{\int_{\caV}  \phi_{\alpha\beta}(\vec X, \vec r, \omega) T(\vec r) \, \diff^3 r}{\int_{\caV}  \phi_{\alpha\beta}(\vec X, \vec r, \omega)\, \diff^3 r}.
\end{equation}

\bibliography{short_paper}
\bibliographystyle{apsrev4-1}